\documentclass[aps,prd,nofootinbib,reprint]{revtex4-2}
\pdfoutput=1

\usepackage[colorlinks=true, urlcolor=blue, citecolor=blue]{hyperref}
\usepackage{graphicx}
\usepackage{amsmath}
\usepackage{physics}
\usepackage{cleveref}
\usepackage{hyperref}
\hypersetup{
    colorlinks=true,
    linkcolor=blue,
    citecolor=blue,
    urlcolor=blue
}

\newcommand{\bubsep}{R_{\text{sep}}} 
\newcommand{\bubrad}{\bar{R}} 

\newcommand{\rhotot}{\rho_\text{tot}}
\newcommand{\rhogs}{\rho_\text{gs}}
\renewcommand{\vec}[1]{\boldsymbol{#1}}
\newcommand{\field}{\vec{\phi}}
\newcommand{\matches}{}

\newcommand{\twolinepower}[2]{\vtop{\hbox{\strut#1} \hbox{\strut$(\! \times \! 10^{#2})$}}}

\newcommand{\betaR}{\beta}
\newcommand{\betaV}{\beta_V}
\newcommand{\vcj}{v_\text{CJ}}

\newcommand{\TS}{\texttt{\allowbreak Trans\-ition\-Sol\-ver}}
\newcommand{\PT}{\texttt{\allowbreak Phase\-Tra\-cer}}
\newcommand{\CT}{\texttt{\allowbreak Cosmo\-Trans\-itions}}

\newcommand{\pt}{\bar{\theta}}
\newcommand{\alpt}{\alpha_{\pt}}

\begin{document}

\title{How robust are gravitational wave predictions from cosmological phase transitions?}

\author{Peter Athron}
\email{peter.athron@njnu.edu.cn}
\affiliation{Department of Physics and Institute of Theoretical Physics, Nanjing Normal University, Nanjing, 210023, China}

\author{Lachlan Morris}
\email{Lachlan.Morris@monash.edu}
\affiliation{School of Physics and Astronomy, Monash University, Melbourne, Victoria 3800, Australia}

\author{Zhongxiu Xu}
\email{211002044@njnu.edu.cn}
\affiliation{Department of Physics and Institute of Theoretical Physics, Nanjing Normal University, Nanjing, 210023, China}

\begin{abstract}
Gravitational wave (GW) predictions of cosmological phase transitions are almost invariably evaluated at either the nucleation or percolation temperature. We investigate the effect of the transition temperature choice on GW predictions, for phase transitions with weak, intermediate and strong supercooling. We find that the peak amplitude of the GW signal varies by a factor of a few for weakly supercooled phase transitions, and by an order of magnitude for strongly supercooled phase transitions. The variation in amplitude for even weakly supercooled phase transitions can be several orders of magnitude if one uses the mean bubble separation, while the variation is milder if one uses the mean bubble radius instead. We also investigate the impact of various approximations used in GW predictions. Many of these approximations introduce at least a 10\% error in the GW signal, with others introducing an error of over an order of magnitude.
\end{abstract}

\maketitle

\section{Introduction}
We are now in an era where existing gravitational wave (GW) data can
have an impact on our understanding of physics beyond the Standard
Model (BSM) of particle physics.  Very recently pulsar timing array experiments have
detected a stochastic GW background (SGWB)~\cite{NANOGrav:2023gor,Xu:2023wog,Antoniadis:2023rey,Reardon:2023gzh}
and find that new physics explanations have a slight preference over
less exotic sources~\cite{NANOGrav:2023hvm}. Existing data on GWs from
the LIGO/VIRGO network~\cite{KAGRA:2023pio} is also constraining
well-motivated Pati-Salam models that can lead to gauge coupling
unification~\cite{Athron:2023aqe} as well as models of the dark sector~\cite{Huang:2021rrk}.

However, with this exciting progress also comes significant challenges.
It is now essential that we have reliable calculations of the
GW spectra for BSM models where we understand the
uncertainties involved and the effects of various approximations and
assumptions that are commonly used.  There are many challenging
calculations involved in going from a particular BSM scenario
to a predicted GW spectrum; see Ref.~\cite{Athron:2023xlk} for a
review. Quantities derived from the effective potential
can strongly depend on the method used~\cite{Athron:2022jyi} and
uncertainties in the GW spectra from effective
potential computations have been investigated in
Ref.~\cite{Croon:2020cgk}.  Here we show that even if the effective
potential calculation was under full control, there are many other
challenges for reliable predictions of GW spectra.

Since the first direct detection of GWs~\cite{LIGOScientific:2016aoc} in 2015,
there has been substantial progress in understanding how to
characterise phase transitions and extract GW predictions. Here we
mention a few important points. Sound waves are expected to be
the largest source of GWs following Ref.~\cite{Hindmarsh:2013xza}
which showed that sound waves source last long after the bubbles have
merged. However, more recently it has been shown that in many cases the
lifetime is nonetheless significantly shorter than the Hubble time~\cite{Ellis:2018mja,Ellis:2020awk} and suppression factors were
introduced~\cite{Ellis:2019oqb,Caprini:2019egz} to account for the
finite lifetime of the source.  These suppression factors were
subsequently refined to address issues stemming from the derivation of
the Hubble time as the maximum lifetime of the source~\cite{Guo:2020grp}.  Furthermore, the modelling of GWs from sound
waves has improved considerably from
simulations~\cite{Hindmarsh:2017gnf, Cutting:2019zws} and the
construction of the sound shell model~\cite{Hindmarsh:2016lnk} and its
further development~\cite{Hindmarsh:2019phv, Wang:2021dwl,
  Cai:2023guc}.  Significant improvements have also been made in determining the kinetic energy fraction that
is available to source GWs. New
parameterisations have been developed that go beyond simplified models such
as the bag model, first for the
case where bubbles expand as supersonic detonations~\cite{Giese:2020rtr} and
later generalised to cover subsonic deflagrations and hybrids~\cite{Giese:2020znk}.
These advances have both improved
predictions and raised questions about our previous and current
understanding of how sensitive GW experiments can be to first-order
phase transitions.

In particular, strongly supercooled phase transitions present
significant challenges for calculations and may lead to erroneous
explanations of GW signals~\cite{Athron:2023mer}.  We therefore treat the
extent of supercooling as an important parameter when considering the
uncertainties and compare scenarios with weak, intermediate, and strong supercooling.
Previously, we have shown that in the presence of supercooling various
possible choices of transition temperature decouple~\cite{Athron:2022mmm}
and it has been argued that the percolation
temperature should be used~\cite{Caprini:2019egz, Wang:2020jrd, Guo:2021qcq, Athron:2022mmm}. Here we show
explicitly that the peak amplitude and frequency of the GW spectrum --- and thus the resulting signal-to-noise ratio (SNR) at a detector --- are sensitive to the choice of transition
temperature. This is especially true for strongly supercooled phase transitions as one might expect, but is also true for weakly supercooled phase transitions. We show
that if one chooses the nucleation temperature as the transition
temperature (as is very common practice), then the peak amplitude, peak frequency, and SNR can change by orders of magnitude compared to when using the percolation temperature. This has a
huge impact on the prospects for detection. However, such a drastic change only arises when using the mean bubble separation as the characteristic length scale. If one is more careful about the choice of length scale, the discrepancy can potentially be reduced to a factor of a few.

Additionally, we investigate how the predictions can be affected by
different estimates of the thermal parameters which determine the
GW spectrum.
We compare various parameterisations of the kinetic energy fraction, which determines the energy available for sourcing GWs.
Another important factor that determines the
peak GW amplitude and frequency is the timescale during which the
source is active, which is usually replaced by a characteristic length
scale. The mean bubble separation is used as this length scale in lattice
simulations. We compare the impact different estimates of this
have on GW signals, and we qualitatively explore the consequences of using the mean bubble radius instead.
Finally, because the
turbulence contribution to the overall GW signal is not well modelled, but could be
significant, we also compare several different choices for the energy
available for sourcing GWs from turbulence and show
the impact that this can have on the SNR.

In \cref{sec:FOPTs_supercooling} we describe first-order phase
transitions and supercooling in more detail, and we define
important milestone temperatures.  In \cref{sec:properties} we
describe how properties of the phase transition and the thermal
parameters are computed in particle physics models. We also
discuss various estimates for these thermal parameters that are made in the literature. We
briefly describe how we use these thermal parameters to predict GW spectra in
\cref{sec:GWs}. We then introduce the model we use to obtain a first-order phase transition in \cref{sec:model}.
Finally, we present our results in \cref{sec:results} and provide concluding remarks in \cref{sec:discussion}.

\section{First-order phase transitions and supercooling}
\label{sec:FOPTs_supercooling}
As the Universe cools down the shape of the effective potential
changes such that minima (or phases) can appear and disappear and cosmological
phase transitions take place. These cosmological phase transitions
play an important role in particle physics, such as breaking the
electroweak symmetry and thereby generating masses for the fundamental
particles via the Higgs mechanism. Further, if a phase transition is of
first order (i.e.\ there is a potential barrier separating the phases),
GWs are produced in the process.

A potential barrier between the phases prevents an instantaneous transition
from the local metastable minimum to the deeper minimum on the other side of the
barrier.  Instead, the phase transition must proceed via
either tunnelling through the barrier or fluctuating over it.  This
first becomes possible when the Universe cools below the critical
temperature, $T_c$, where the free energy densities of the two minima are degenerate.  Below $T_c$
the transition begins through a stochastic process where
the tunnelling or fluctuations occur at localised points in spacetime,
and when this happens bubbles of the new phase can form and grow in
a process known as bubble nucleation.  The phase transition completes if
the bubbles of the new phase fill the whole universe.  More precisely,
because it is a stochastic process we define the completion
temperature, $T_f$, to be the temperature when the fraction of the
universe left in the false vacuum (i.e.\ the old phase) is less then $1\%$, $P_f(T_f) <
0.01$.

When this process takes a long time to complete $T_f$ may be much
smaller than the critical temperature $T_c$ at which the new minimum
first becomes energetically favoured.  This is known as supercooling
in analogy with the phenomenon where liquids are supercooled well below
their freezing point.  All first-order cosmological phase transitions exhibit some
degree of supercooling because they do not happen instantly. However, the
temperature change can vary from $T_f$ being within 1\%
of $T_c$ to being orders of magnitude smaller.

The degree of supercooling can have a significant impact on a phase
transition and is an important characteristic when comparing phase
transitions. Increasing supercooling may boost the energy released in
the phase transition and the amplitude of
resultant GWs, but too much supercooling can
lead to the transition failing to complete.

Strongly supercooled phase transitions admit qualitatively different
behaviour compared to weakly supercooled phase transitions.  Because
the nucleation rate is lower, the smaller number of bubbles that are
nucleated grow to much larger sizes. This means that the number of
bubbles per Hubble volume, $N$, can be less than one during the period
where most of the bubbles are colliding or even by the time the phase
transition has completed~\cite{Athron:2022mmm}. This can be expressed
more precisely as follows. The nucleation temperature $T_n$ is defined
by the condition $N(T_n) = 1$. Usually $T_n$ is higher than the
percolation temperature $T_p$, defined by the moment when the false
vacuum fraction, $P_f$, is roughly 71\%: $P_f(T_p)=0.71$. Roughly
speaking, $T_p$ is where the bubbles should be in contact with each
other (see section 4.7.2 of Ref.~\cite{Athron:2023xlk} for more
details). In strongly supercooled scenarios the nucleation temperature
can be reached some time after most of the bubble collisions have
taken place.  In more extreme cases the phase transition may complete,
reaching $P_f(T_f) < 0.01$, before $N(T) = 1$. In such cases there is
no nucleation temperature.  However, strongly supercooled scenarios
can also have enough bubble nucleation such that $N(T) = 1$ is reached
relatively early in the phase transition but the transition is still
slow leading to a substantial gap between $T_n$ and $T_p$ or $T_f$.
Thus, the nucleation temperature is not coupled with the actual
progress of the phase transition and the production of GWs.

\section{Determining properties of the phase transition}
\label{sec:properties}
\subsection{Temperatures and length scales} \label{sec:temps-and-lens}
The rate of a phase transition depends strongly on the size and persistence of the potential barrier. In fast
transitions the barrier disappears fairly quickly. The nucleation rate
is initially zero at $T_c$ and then increases
rapidly as the barrier dissolves, giving an exponential nucleation
rate of the form
\begin{align}
	\Gamma(t) = \Gamma(t_*) \exp(\betaR(t-t_*)),\label{Eq:expoNucRate}
\end{align}
where $t_*$
is some relevant time in the transition (often taken to correspond to $T_n$). In contrast, if the
barrier persists at low temperatures or even at $T=0$, the nucleation
rate can instead reach a maximum at some temperature $T_\Gamma$ because
lower temperature reduces the likelihood of thermal fluctuations over the barrier.

The nucleation rate is given by~\cite{Linde:1981zj}
\begin{align}
  \Gamma(T) = T^4 \left(\frac{S(T)}{2\pi}\right)^{\!\!\frac32}\exp(-S(T)),
  \label{Eq:NucRate}
 \end{align} 
where $S(T)$ is the bounce action which we obtain from a modified
version of \CT{}~\cite{Wainwright:2011kj}.%
\footnote{See appendix F of Ref.~\cite{Athron:2022mmm} for details of the modifications.} Issues related to the use of \cref{Eq:NucRate} are discussed and addressed in Refs.\ \cite{Gould:2021ccf,Hirvonen:2021zej,Lofgren:2021ogg,Ekstedt:2023sqc}.  Here we investigate issues that arise even if these issues with the nucleation rate and those of the effective potential (which are strongly related) are completely accounted for. If one expresses $S$ as a function of time and Taylor
expands about $t_*$,
\begin{align}
  S(t) \approx S(t_*) & + \left.\dv{S}{t}\right|_{t=t_*} \!\!\!\! (t-t_*) \\
  & {} + {} \frac12 \! \left.\dv[2]{S}{t}\right|_{t=t_*} \!\!\!\! (t-t_*)^2 + \cdots ,
 \label{Eq:talyorexpandbounce}
\end{align}
then truncating at first order gives the exponential nucleation rate
given in \cref{Eq:expoNucRate}, and we can identify
\begin{align}
  \betaR \equiv - \! \left.\dv{S}{t}\right|_{t=t_*} . 
  \label{Eq:beta}
\end{align}
This can be useful because $\betaR$ is related to the mean separation of
bubbles, $\bubsep$, through~\cite{Enqvist:1991xw}
\begin{equation}
  \bubsep = (8\pi)^{\frac13} \frac{v_w}{\betaR}.
  \label{Eq:Rsep_from_beta}
\end{equation}
The mean bubble separation is an important quantity for GW predictions. \Cref{Eq:Rsep_from_beta} should hold when evaluated at the temperature where $P_f$ has decreased to $1/e$, denoted by $T_e$. Computing $\betaR$ directly from the bounce action and
using \cref{Eq:Rsep_from_beta} to estimate $\bubsep$ can simplify calculations significantly.

However, while an exponential nucleation rate is a common assumption
and \cref{Eq:Rsep_from_beta} is widely used, these approximations can be
problematic in strongly supercooled scenarios. We will demonstrate the
potential consequences of this in \cref{sec:results}. Note that if the transition temperature $T_*$ used to evaluate
$\betaR$ is close to the temperature where nucleation rate is maximised,
$T_\Gamma$, then $\betaR \approx 0$. Further, $\betaR$ is negative when $T_* < T_\Gamma$.
Therefore, the use of $\betaR$
entirely breaks down in these cases.  However, because $\betaR$ vanishes
one can truncate \cref{Eq:talyorexpandbounce} at second order and obtain a Gaussian
nucleation rate,
\begin{align}
	\Gamma(t) = \Gamma(t_*) \exp(-\frac{\betaV^2}{2}(t-t_*)^2),\label{Eq:GaussianNucRate}
\end{align}
where
\begin{align}
  \betaV =  \sqrt{\left.\dv[2]{S}{t}\right|_{t=t_\Gamma}}.
  \label{Eq:betaV}
\end{align}
We can relate $\betaV$ to $\bubsep$ through~\cite{Ellis:2018mja}
 \begin{equation}
   \bubsep = \left(\sqrt{2 \pi} \frac{\Gamma(T_\Gamma)}{\betaV} \right)^{\!\!-\frac13} .
 \label{Eq:bubsep_from_betav}  
\end{equation}
It is unclear how well the approximations \cref{Eq:Rsep_from_beta} and \cref{Eq:bubsep_from_betav} perform, so we include this investigation in our study. We note that we use temperature rather than time in our analysis, so we employ the usual time-temperature relation~\cite{Athron:2022mmm}
\begin{equation}
	\dv{t}{T} = \frac{-1}{T H(T)} .
\end{equation}
Thus, $\betaR$ and $\betaV$ are in fact calculated from $\text{d}S/\text{d}T$. The Hubble rate is given by
\begin{align}
	H(T) =\sqrt{\frac{8\pi G}{3} \rhotot(T)} ,
\end{align}
where $\rhotot$ is the total energy density. We use energy conservation such that $\rhotot = \rho_f - \rhogs$, where $\rho_f$ is the false vacuum energy density and $\rhogs$ is the ground state energy density. We renormalise the free energy density such that $\rhogs = 0$, leaving $\rhotot = \rho_f$.

Returning to the full treatment, the nucleation rate in
\cref{Eq:NucRate} can be used directly to compute the false vacuum
fraction $P_f$ as a function of temperature, given by
\begin{equation}
 P_f(T) = \exp\!\left[-\frac{4\pi}{3} \! \int_T^{T_c} \! \frac{dT'}{T'^4} \frac{\Gamma(T')}{H(T')} \! \left(\int_T^{T'} \!\!\! dT'' \frac{v_w(T'')}{H(T'')} \right)^{\!\!3}\right] . \label{Eq:false_vac_frac}
\end{equation} 
Here we have assumed that the Universe is expanding adiabatically and we neglect the initial radius of the bubble at formation.  See Ref.~\cite{Athron:2023xlk} for more details on the derivations and assumptions. The last undetermined quantity in \cref{Eq:false_vac_frac} is the bubble wall velocity, $v_w$. We discuss our treatment of $v_w$ in \cref{sec:hydro-params}.

The number of bubbles nucleated at any given temperature can also be computed from \cref{Eq:NucRate}. In the literature it is standard to calculate the nucleation temperature from an approximation for the number of bubbles per Hubble volume,
\begin{align}
 N(T) &=  \int_{T}^{T_c} \!\! dT' \, \frac{\Gamma(T') }{T'H^4(T')}. \label{Eq:Approx_Number_bub_per_hub_vol}
\end{align}
This implicitly assumes a fast transition so that one can assume $P_f = 1$ before $T_n$, and thus omit $P_f$ from the integrand~\cite{Athron:2022mmm}.\footnote{A factor of $4\pi/3$
  from the spherical Hubble volume is also neglected in this
  treatment.}
In this study we only use $T_n$ to
show the impact of approximations made in the literature, so we use the
expression in \cref{Eq:Approx_Number_bub_per_hub_vol} to calculate
$T_n$ for consistency.

In contrast, to compute the mean bubble separation we determine the
bubble number density with $P_f(T)$ included to account for the fact
that true vacuum bubbles can only nucleate in regions that are
still in the false vacuum. The mean bubble separation is given by
\begin{equation}
	\bubsep(T) = (n_B(T))^{-\frac13} , \label{Eq:bubsep_from_bnd}
\end{equation}
where
\begin{equation}
	n_B(T) = T^3 \!\! \int_T^{T_c} \! dT' \frac{\Gamma(T') P_f(T')}{T'^4 H(T')} \label{Eq:bubble_num_density}
\end{equation}
is the bubble number density. Finally, there are possibly other choices for the characteristic
length scale in GW predictions~\cite{Megevand:2016lpr, Cai:2017tmh, Ellis:2018mja, Wang:2020jrd, Athron:2023xlk}. However, fits for GW predictions are determined in terms of $\bubsep$, and one cannot directly replace $\bubsep$ with alternative length scales in those fits. Still, we seek to investigate (among other things) the impact of the choice of $T_*$ on the GW predictions (see \cref{sec:results}), so it is important to understand the impact of $T_*$ on various length scales. Thus, we also consider the mean bubble
radius,
\begin{equation}
	\bubrad(T) = \frac{T^2}{n_B(T)} \int_T^{T_c} \!\! dT' \frac{\Gamma(T') P_f(T')}{T'^4 H(T')} \! \int_T^{T'} \!\! dT'' \frac{v_w(T'')}{H(T'')} .
\label{Eq:meanBubbleRadius}
\end{equation}
For more details see section 5.5 of Ref.~\cite{Athron:2023xlk}
and references therein.

We can compute the milestone temperatures $T_n$, $T_p$, $T_e$ and $T_f$ using \cref{Eq:Approx_Number_bub_per_hub_vol,Eq:false_vac_frac}, and we can similarly use \cref{Eq:bubsep_from_bnd,Eq:meanBubbleRadius} to compute $\bubsep$ and $\bubrad$ at these milestone temperatures or at arbitrary temperatures. We use \PT{}~\cite{Athron:2020sbe} to map the phase structure of the potential and \TS{}~\cite{Athron:2023ts} to analyse the phase history, including all relevant phase transitions,%
\footnote{There is another high-temperature phase transition with $T_c \sim 180$ GeV in the intermediate and strong supercooling scenarios considered in \cref{sec:model}. The phase transition is very fast and is not relevant to our analysis.}
as well as determine the milestone temperatures and relevant parameters for GW predictions.
The GW fits are parameterised in terms of thermal parameters, which --- in
addition to the transition temperature and the characteristic length scale --- also include hydrodynamic parameters such as the kinetic
energy fraction and the bubble wall velocity.

\subsection{Hydrodynamic parameters} \label{sec:hydro-params}

Here we discuss the hydrodynamic parameters used in GW fits. First we discuss our best treatment for these parameters, then we introduce several common variations to this treatment used in the literature. We will investigate the impact of these variations on the GW signature in \cref{sec:results-treatment}. All of these parameters --- and all of the quantities that they depend on --- should be evaluated at the transition temperature, $T_*$.

In our best treatment, we take $T_* = T_p$, and we determine the kinetic energy fraction using the pseudotrace difference between the phases, corresponding to M2 in Ref.~\cite{Giese:2020rtr}:
\begin{equation}
	K = \frac{\pt_f(T_*) - \pt_t(T_*)}{\rhotot(T_*)} \kappa_{\pt}(\alpt(T_*), c_{s,f}(T_*)).\label{Eq:KEfrac}
\end{equation}
Here, $c_{s,f}$ is the speed of sound in the false vacuum and $\alpt$ is the transition strength parameter. The speed of sound in phase $i$ is given by \cite{Athron:2023xlk}
\begin{equation}
  c_{s,i}^{2}(T) = \left.\frac{\partial_T V}{T \partial_T^2 V}\right\vert_{\field_i(T)} ,
  \label{Eq:speedofsound}
\end{equation}
where $V$ is the effective potential, or free energy density, and
$\field_i$ is the field configuration for phase $i$. The transition
strength parameter is defined as
\begin{equation}
	\alpha_x(T) = \frac{4 \! \left(x_f(T) - x_t(T)\right)}{3 w_f(T)} , \label{eq:alpha}
\end{equation}
where $x$ is a hydrodynamic quantity for which various choices exist in the literature, and $w_f$ is the enthalpy density in the false vacuum. We use the pseudotrace for $x$ in our best treatment, given by~\cite{Giese:2020rtr}
\begin{equation}
	\pt_i(T) = \frac14 \! \left(\rho_i(T) - \frac{p_i(T)}{c_{s,t}^{2}(T)} \right) 
\end{equation}
in phase $i$, where $\rho$ and $p$ are the energy density and pressure, respectively. The pseudotrace generalises the trace anomaly to models where the speed of sound deviates from $1/\sqrt{3}$. We use the code snippet provided in the appendix of Ref.~\cite{Giese:2020rtr} to determine the efficiency coefficient $\kappa_{\pt}$.  The impact of variations in the speed of sound and other approximations have been previously considered in Refs.\ \cite{Giese:2020znk, Wang:2020nzm, Wang:2021dwl, Tenkanen:2022tly}.

Turbulence from cosmological phase transitions is not well understood because current hydrodynamic simulations cannot probe the turbulent regime. Hence, it is difficult to estimate the efficiency coefficient for turbulence, $\kappa_\text{turb}$, which is needed for turbulence contributions to the production of GWs. However, it is expected that stronger phase transitions (with larger $\alpha$) could result in more turbulence developing sooner and could reduce the lifetime of the sound wave source. Lacking sufficient modelling of the turbulence source, we consider the efficiency coefficient as a fraction of $\kappa_{\pt}$,
\begin{equation}
	\kappa_\text{turb} = \epsilon \kappa_{\pt} , \label{Eq:epsilon-turb} 
\end{equation}
and we take $\epsilon = 0.05$ as our default treatment.

Finally, for our treatment of the bubble wall velocity, we assume
bubbles grow as supersonic detonations regardless of the extent of
supercooling for simplicity. General friction estimates are beyond the
scope of this study, and neither the ultra-relativistic or
non-relativistic limits of friction are applicable for all benchmark points in our
study. We assume the bubbles expand at the Chapman-Jouguet velocity,
\begin{equation}
	v_w = \vcj = \frac{1 + \sqrt{3\alpt (1 + c_{s,f}^2 (3\alpt - 1))}}{c_{s,f}^{-1} + 3 \alpt c_{s,f}} , \label{eq:vcj}
\end{equation}
where temperature dependence has been suppressed.
The Chapman-Jouguet velocity is by no means the most likely supersonic
detonation solution, however it does capture dependence on the
transition temperature and ensures a supersonic detonation regardless
of the extent of supercooling. The same cannot be said for any fixed
choice of $v_w$.

We now turn to the variations on our best treatment. First, we consider the effect of setting $T_*$ to the other milestone temperatures: $T_n$, $T_e$ and $T_f$. This involves using our best treatment (e.g.\ calculating $K$ using \cref{Eq:KEfrac}) but evaluating all quantities at, for example, $T_n$ instead of $T_p$. As a reminder, $T_n$ can be obtained by the condition $N(T_n) = 1$ (see \cref{Eq:Approx_Number_bub_per_hub_vol}), while $T_p$, $T_e$ and $T_f$ all come from conditions on the false vacuum fraction (see \cref{Eq:false_vac_frac}); specifically, $P_f(T_p) = 0.71$, $P_f(T_e) = 1/e$ and $P_f(T_f) = 0.01$.

The approach we use for estimating $K$ was developed only recently in Refs.~\cite{Giese:2020rtr, Giese:2020znk}, so it is not yet widely adopted.
More approximate treatments are widespread, which we enumerate here. It is very common to determine $K$ through
\begin{equation}
  K_{x}=\frac{\kappa_{x} \alpha_{x}}{1+\alpha_{x}} ,
  \label{Eq:Kalpha}
\end{equation}
with various choice of $x$ often being made. This parameterisation alone introduces error in the determination of $K$, regardless of the choice of $x$ (see \cref{app:deltas} for details). The trace anomaly,
\begin{equation}
	\theta(T) = \frac14 (\rho(T) - 3p(T)) ,
\end{equation}
is the closest alternative to $\pt$, in fact exactly matching $\pt$ when $c_{s,t} = 1/\sqrt{3}$ like in the bag model. The other common choices for $x$ are the pressure $p$ and the energy density $\rho$. The efficiency coefficient used for these choices of $x$ was derived in the bag model, and is given by~\cite{Espinosa:2010hh}
\begin{equation}
	\kappa = \frac{\sqrt{\alpha_x}}{0.135 +\sqrt{0.98 + \alpha_x}} \label{eq:kappa-vcj}
\end{equation}
for $v_w = \vcj$, which is implicitly dependent on temperature.

In these more approximate treatments of $K$, the enthalpy density in the denominator of
\cref{eq:alpha} is usually replaced with $w_{f} = \frac{4}{3}
\rho_{R}$, where $\rho_{R}=\frac{\pi^2}{30}g_\text{eff} T^4$ is the radiation energy density
and $g_\text{eff}$ is the effective number of relativistic degrees of freedom. We find the
replacement of the enthalpy density in this way (which comes from the bag
model) to be a very good approximation. This replacement leads to less than 1\% error in the GW predictions.
Therefore our
$\alpha_\rho$ effectively corresponds to the latent heat definition
frequently found in the literature, see eq.~5.35 of
Ref.~\cite{Athron:2023xlk}.  Similarly $\alpha_\theta$ also
effectively corresponds to eq.~5.36 of Ref.~\cite{Athron:2023xlk},
which also frequently appears in the literature, though here one also
needs to substitute $\theta = \frac14(\rho-3p)$. One could also replace $\pt$ with $\theta$ in \cref{Eq:KEfrac} and use \cref{eq:kappa-vcj} for $\kappa$, corresponding to M3 in Ref.~\cite{Giese:2020rtr}. However, we find this introduces at most 1\% difference in the GW predictions compared to using \cref{Eq:Kalpha} with $x = \theta$, so we do not consider this variation in our results.

As described in \cref{sec:temps-and-lens}, one can approximate the mean bubble separation $\bubsep$ through the often-used thermal parameter $\betaR$, or through $\betaV$. We investigate the error in these approximations for $\bubsep$ and the corresponding effect on GW predictions. We also demonstrate the impact of using $\bubrad$ instead of $\bubsep$, but we do not treat this as a variation of the treatment because mapping $\bubrad$ onto existing GW fits is currently ambiguous.

We also consider alternative treatments of the turbulence efficiency coefficient. The most obvious variation is to simply choose another arbitrary, fixed value. We choose $\epsilon_2 = 0.1$, where the subscript `2' denotes the index of this variation for $\epsilon$. We also consider $\epsilon_3 = \left(1 - \min(H(T_*) \tau_\text{sw}, 1\right))^{2/3}$, which comes from assuming that a reduction in the lifetime of the sound wave source $\tau_\text{sw}$ could boost the turbulence contribution to GW production~\cite{Alanne:2019bsm, Ellis:2019oqb}. However, the effective lifetime of the sound wave source is more accurately suppressed by the factor $\Upsilon = 1 - 1/\sqrt{1 + 2 H(T_*) \tau_\text{sw}}$ derived in Ref.~\cite{Guo:2020grp}. This motivates a slightly modified choice: $\epsilon_4 = (1-\Upsilon)^{2/3}$.

There are of course many other variations to the treatment that could be considered, but we restrict our study to the variations mentioned thus far. Changes to the bubble wall velocity could significantly impact the GW predictions and even the phase transition properties, particularly if the expansion mode of the bubbles changes from a supersonic detonation. \TS{} currently does not use a full hydrodynamic treatment of bubble profiles and therefore only provides accurate predictions for supersonic detonations.%
\footnote{Reheating in the false vacuum for other bubble expansion modes affects both bubble nucleation and growth~\cite{Heckler:1994uu, Megevand:2017vtb, Ajmi:2022nmq}.}
Thus, we currently cannot explore effect of $v_w$ on GW predictions. We explored the impact of approximations made for the reheating temperature and GW redshifting factors in Ref.~\cite{Athron:2023mer}, and found that their effects were small. We do not reconsider these approximations here due to their accuracy. Also, we explored accuracy of various approximations for $T_n$ as a function of supercooling in Ref.~\cite{Athron:2022mmm}. Here we only calculate $T_n$ using \cref{Eq:Approx_Number_bub_per_hub_vol}, but we note that rougher approximations for $T_n$ are unreliable in strongly supercooled scenarios, and would thus lead to significant errors in GW predictions.

\section{Gravitational waves} \label{sec:GWs}

We consider only the sound wave and turbulence sources of GWs in this study. The collision source is expected to contribute negligibly due to friction with the plasma. Even though some of the benchmark points listed in \cref{sec:model} admit strong supercooling, the bubbles nucleate at temperatures where the plasma still imposes significant friction on the expanding bubble walls. Thus, we do not expect runaway bubble walls and consequently neglect the collision source altogether.  

The general scaling of the GW equations is predominantly governed by two key parameters: the kinetic energy fraction $K$ and the characteristic length scale $L_*$. We set $L_* = \bubsep(T_p)$ in our best treatment. The scaling of the peak amplitude $\Omega_\text{peak}$ and the peak frequency $f_\text{peak}$ is roughly
\begin{align}
	\Omega_\text{peak} & \propto K^n L_* , \\
	f_\text{peak} & \propto L_*^{-1} ,
\end{align}
where $n = 2$ for sound waves and $n = 3/2$ for turbulence.

The details of the GW equations we use can be found in appendix A.5 of Ref.~\cite{Athron:2023mer}. In addition to the turbulence fit~\cite{Caprini:2009yp} and the sound shell model~\cite{Hindmarsh:2016lnk, Hindmarsh:2019phv} used for the sound wave source, we also consider another fit for the sound wave source provided in Ref.~\cite{Hindmarsh:2017gnf}. We will refer to this fit as the `lattice fit' for the sound wave source, for lack of a better name. In this fit, the redshifted peak amplitude is
\begin{equation}
	h^2 \Omega_\text{sw}^\text{lat}(f) = 5.3 \! \times \! 10^{-2} \, \mathcal{R}_\Omega K^2 \! \left(\!\frac{H L_*}{c_{s,f}}\! \right) \! \Upsilon(\tau_\text{sw}) S_\text{sw}(f) ,
\end{equation}
the redshifted peak frequency is
\begin{equation}
	f_\text{sw}^\text{lat} = 1.58 \, \mathcal{R}_f \! \left(\frac{1}{L_*} \right) \! \left(\frac{z_p}{10} \right) ,
\end{equation}
matching one of the key frequencies in the sound shell model, and the spectral shape is
\begin{equation}
	S_\text{sw}(f) = \left(\frac{f}{f_\text{sw}^\text{lat}} \right)^{\!\!3} \! \left(\frac{7}{4 + 3(f/f_\text{sw}^\text{lat})^2} \right)^{\!\!\frac72} .
\end{equation}
See Ref.~\cite{Athron:2023xlk} and the appendices of Ref.~\cite{Athron:2023mer} for details of the redshifting factors $\mathcal{R}_f$ and $\mathcal{R}_\Omega$, the lifetime suppression factor $\Upsilon$, and the simulation-derived factor $z_p$ (which is taken to be $z_p = 10$). All quantities in the fit are evaluated at $T_*$, except for the redshifting factors. These are instead evaluated at the reheating temperature, which itself depends on $T_*$. Just as in Ref.~\cite{Athron:2023mer}, we do not include a suppression factor coming from bubbles not reaching their asymptotic hydrodynamic profiles in the simulations from which the GW fits are obtained. This suppression factor would likely depend on $T_*$ and the extent of supercooling, however further modelling is required.

We also compute the SNR for the planned space-based GW detector LISA~\cite{LISA:2017pwj}. LISA has a peak sensitivity at the frequency scale $f_\text{LISA} \sim 10^{-3}$ Hz, which is the expected scale of GW signals from a first-order electroweak phase transition~\cite{Caprini:2015zlo}. We use the projected sensitivity curve $\Omega_\text{LISA}$ from Refs.~\cite{Robson:2018ifk, Caprini:2019pxz}, plotted in \cref{fig:gw-peak-scatter}. We calculate the SNR as \cite{Caprini:2019pxz}
\begin{equation}
	\text{SNR} = \sqrt{\mathcal{T} \! \int_0^\infty \!\!\! df \! \left(\frac{\Omega_\text{GW}(f)}{\Omega_\text{LISA}(f)} \right)^{\!\!2}} ,
\end{equation}
where $\Omega_\text{GW}$ is the total GW signal from the sound wave and turbulence sources, and assume an effective observation time $\mathcal{T}$ of three years, coming from a mission duration of four years and 75\% data-taking uptime.

\section{Model} \label{sec:model}
We use the real scalar singlet model --- which is a simple yet realistic extension to the Standard Model --- to realise a first-order electroweak phase transition. Details of this model and our treatment of one-loop corrections are available in section 4.2 of Ref.~\cite{Athron:2022mmm}. We improve the treatment by adding extra fermions (including all quarks and the muon and tau), and adding Boltzmann suppression factors to the Debye corrections. We also appropriately adjust the radiation degrees of freedom to $g_*' = 22.25$. A similar treatment in a simpler single-field model was used in Ref.~\cite{Athron:2023mer}.

We consider four benchmark points (BPs) in this study, each with a different extent of supercooling. All BPs come from a narrow slice of the total parameter space of the model. We start with M2-BP2 of Ref.~\cite{Athron:2022mmm} and vary only the mixing angle $\theta_m$ to vary the extent of supercooling. The other input parameters are fixed as $\kappa_{hhs} = -1259.83$ GeV, $\kappa_{sss} = -272.907$ GeV, $v_s = 663.745$ GeV and $m_s = 351.183$ GeV. The mixing angles and the milestone temperatures for the BPs are listed in \cref{tab:bp}. The supercooling increases with the BP index. BP1 represents a typical weakly supercooled phase transition with only 1 GeV difference between the onset of bubble nucleation and percolation, and $\alpt \approx 0.01$. BP2 represents a moderately supercooled phase transition with $\alpt \approx 0.05$. Both of these BPs have an exponential nucleation rate, thus we do not calculate $T_\Gamma$ for them. BP3 represents a very strongly supercooled phase transition, where the physical volume of the false vacuum only begins to decrease just below $T_p$. While BP3 has a completion temperature, percolation is questionable~\cite{Turner:1992tz, Ellis:2018mja, Athron:2022mmm}. The transition strength parameter is $\alpt \approx 1.7$, beyond the reach of current hydrodynamic simulations of GWs~\cite{Cutting:2019zws}. Thus, one must be cautious when interpreting GW predictions from BP3, and indeed BP4. BP4 has even stronger supercooling, so much so that the phase transition does not complete. The transition strength parameter in BP4 is $\alpt \approx 177$.

\begin{table}
\begin{tabular}{|l|l|l|l|l|l|l|l|l|}
	\hline
	& \multicolumn{1}{|c|}{$\theta_m$} & \multicolumn{1}{|c|}{$T_c$} & \multicolumn{1}{|c|}{$T_n$} & \multicolumn{1}{|c|}{$T_p$} & \multicolumn{1}{|c|}{$T_e$} & \multicolumn{1}{|c|}{$T_f$} & \multicolumn{1}{|c|}{$T_\Gamma$} & \multicolumn{1}{|c|}{$\log_{10}(\alpt)$} \\
	\hline
	BP1 & 0.24 & 117.0 & 106.0 & 104.8 & 104.7 & 104.6 & N/A & $-1.938$ \\
	\hline
	BP2 & 0.258 & 108.3 & 78.10 & 74.17 & 73.80 & 73.24 & N/A & $-1.264$ \\
	\hline
	BP3 & 0.262 & 106.2 & N/A & 32.46 & 25.65 & 12.69 & 59.47 & $\;\;\,0.2178$ \\
	\hline
	BP4 & 0.2623 & 106.1 & N/A & 10.09 & N/A & N/A & 59.57 & $\;\;\,2.248$ \\
	\hline
\end{tabular}
\caption{Benchmark points and their corresponding milestone temperatures. The mixing angle is expressed in radians, and the temperatures have units of GeV. The transition strength parameter $\alpt$ is evaluated at $T_p$.}
\label{tab:bp}
\end{table}

\section{Results} \label{sec:results}

\subsection{Dependence on the transition temperature} \label{sec:results-temperature}

\begin{figure}
	\includegraphics[width=0.95\linewidth]{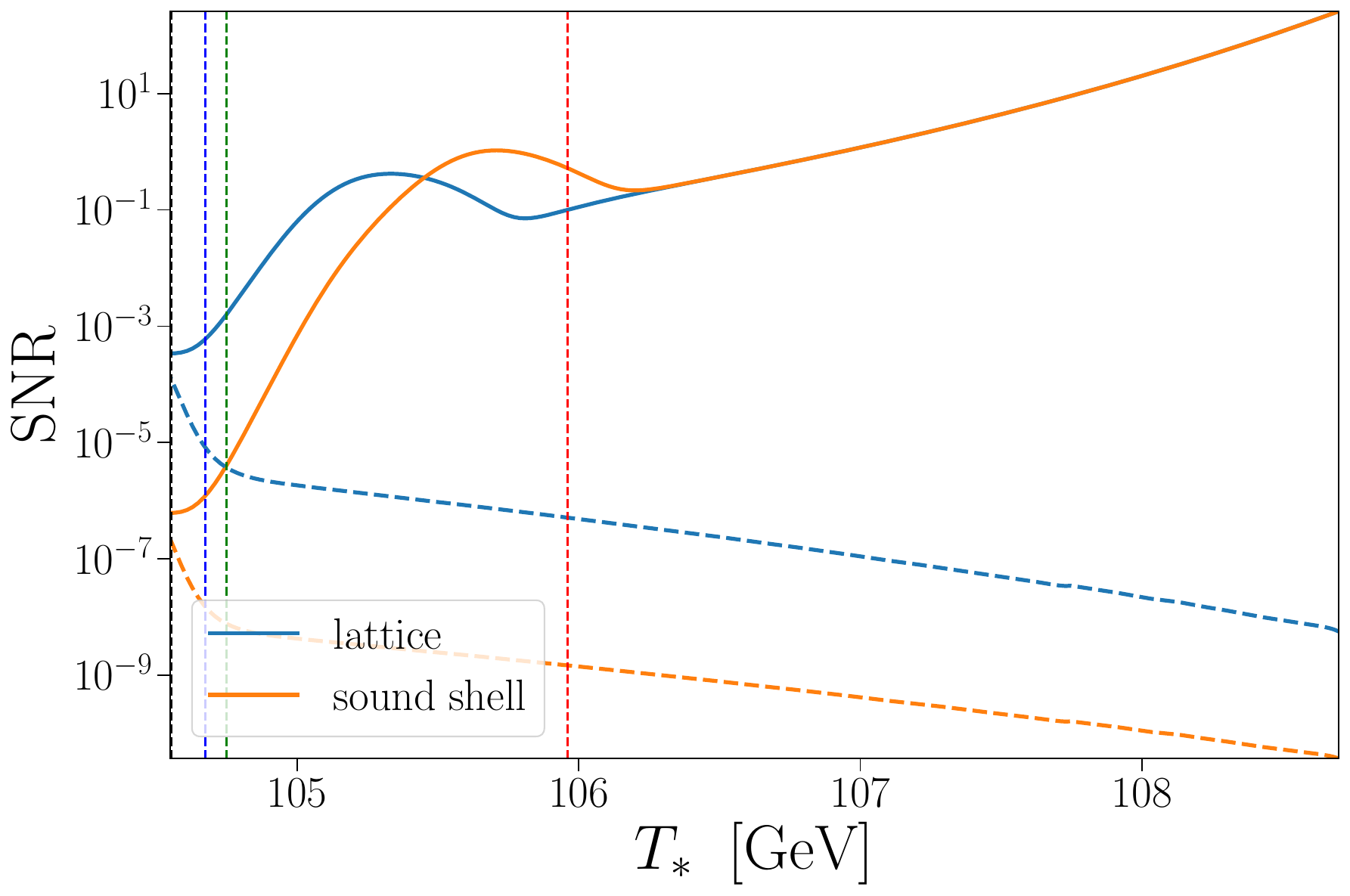}\\
	\includegraphics[width=0.95\linewidth]{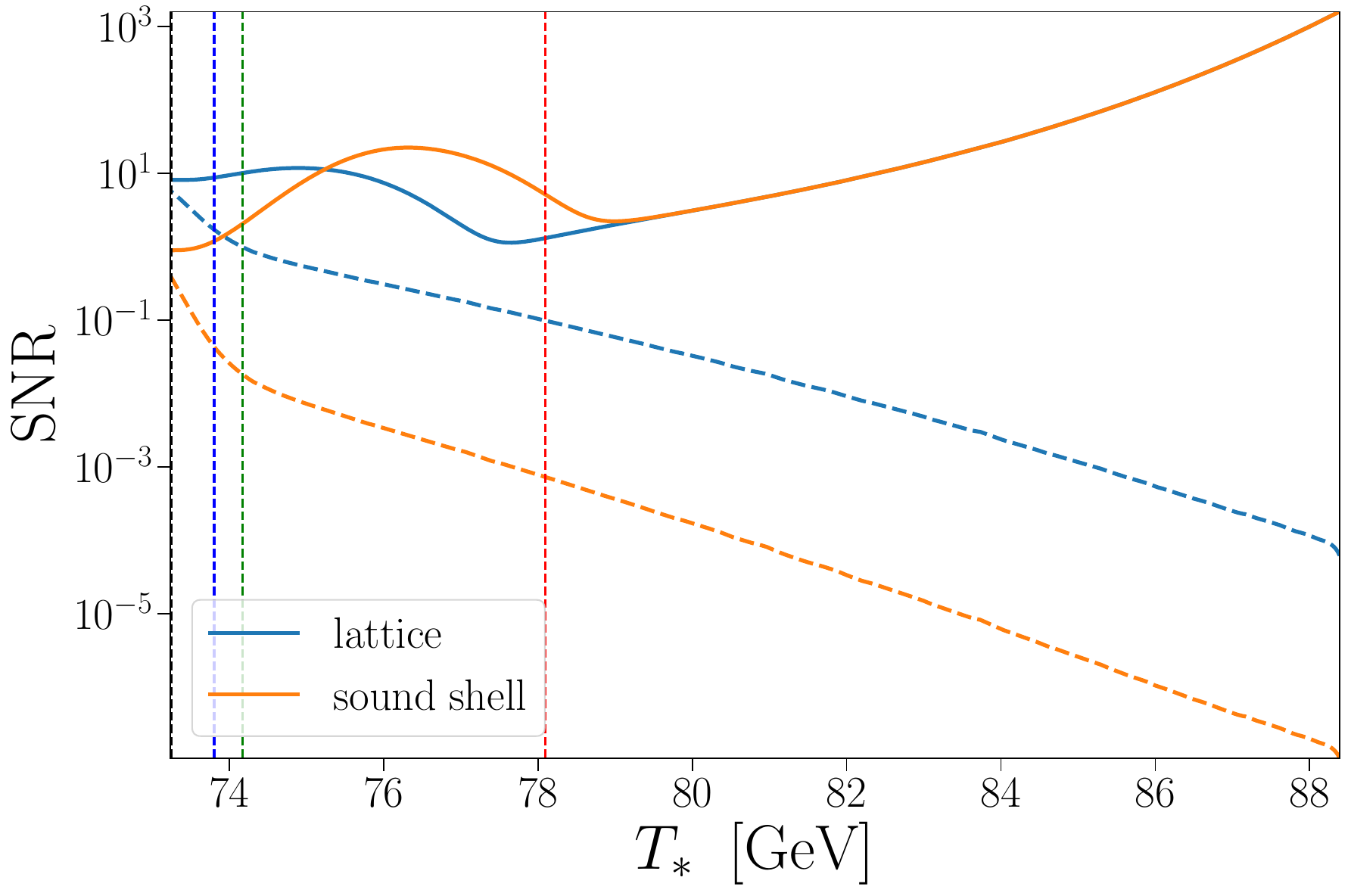}\\
	\includegraphics[width=0.95\linewidth]{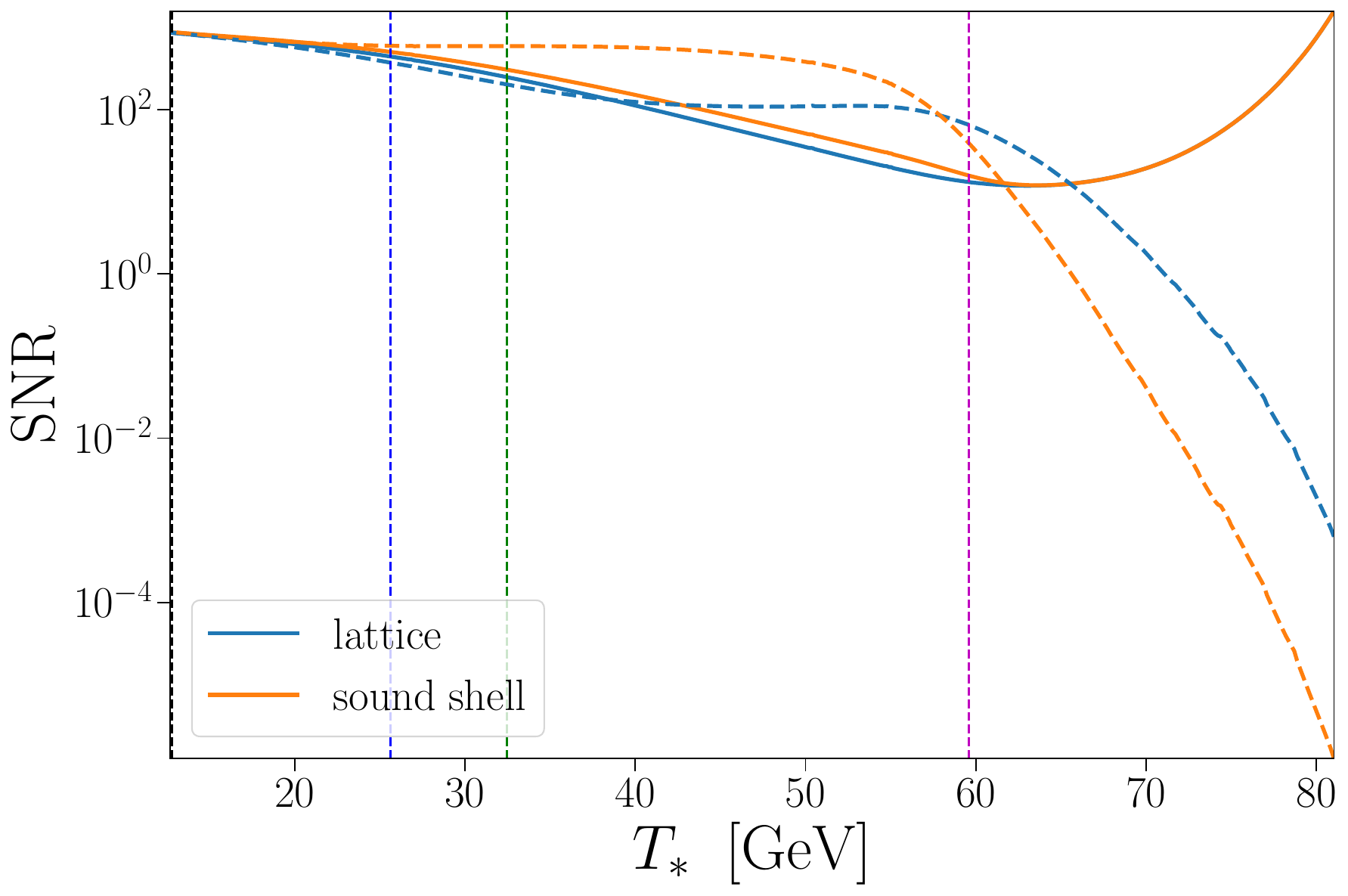}\\
	\includegraphics[width=0.95\linewidth]{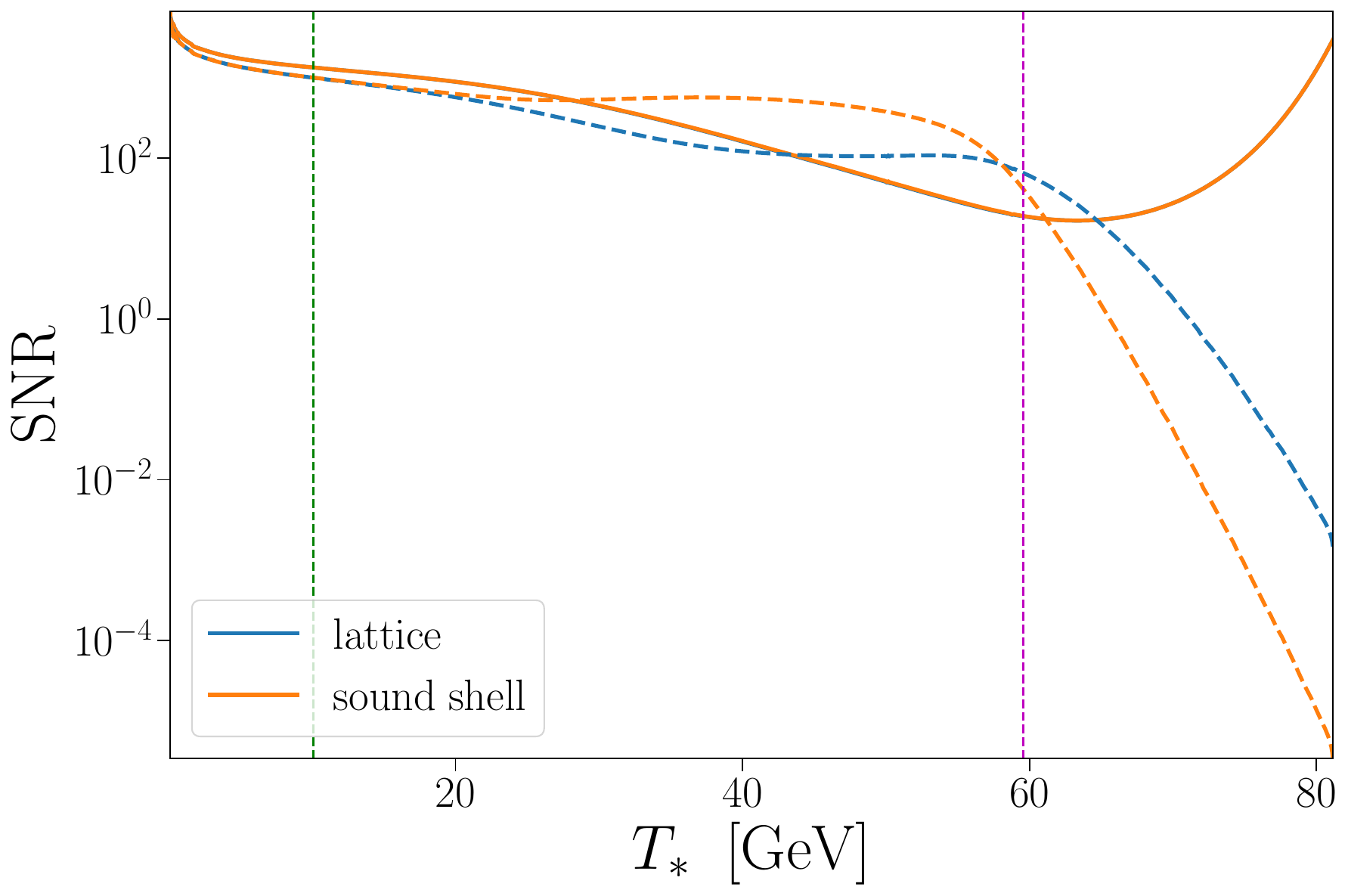}
	\caption{SNR at LISA as a function of $T_*$. From top to bottom: (a) BP1, (b) BP2, (c) BP3, (d) BP4. The vertical dashed lines correspond to key temperatures: $T_\Gamma$ (magenta), $T_n$ (red), $T_p$ (green), $T_e$ (blue) and $T_f$ (black). Completion occurs at the left border of each plot, except for BP4 where there is no completion. The solid curves correspond to $L_* = \bubsep$ and the dashed curves correspond to $L_* = \bubrad$.}
	\label{fig:SNR-vs-T}
\end{figure}

In this section we discuss the impact on GW predictions when varying the transition temperature, $T_*$. The SNR as a function of $T_*$ is shown in \cref{fig:SNR-vs-T} for each BP. The SNR varies by orders of magnitude over the duration of the phase transition. However, GWs are not produced until the phase transition is well underway, so we restrict our attention to the temperature range $T \in [T_f, \max(T_n, T_\Gamma)]$.

\begin{figure}
	\includegraphics[width=0.95\linewidth]{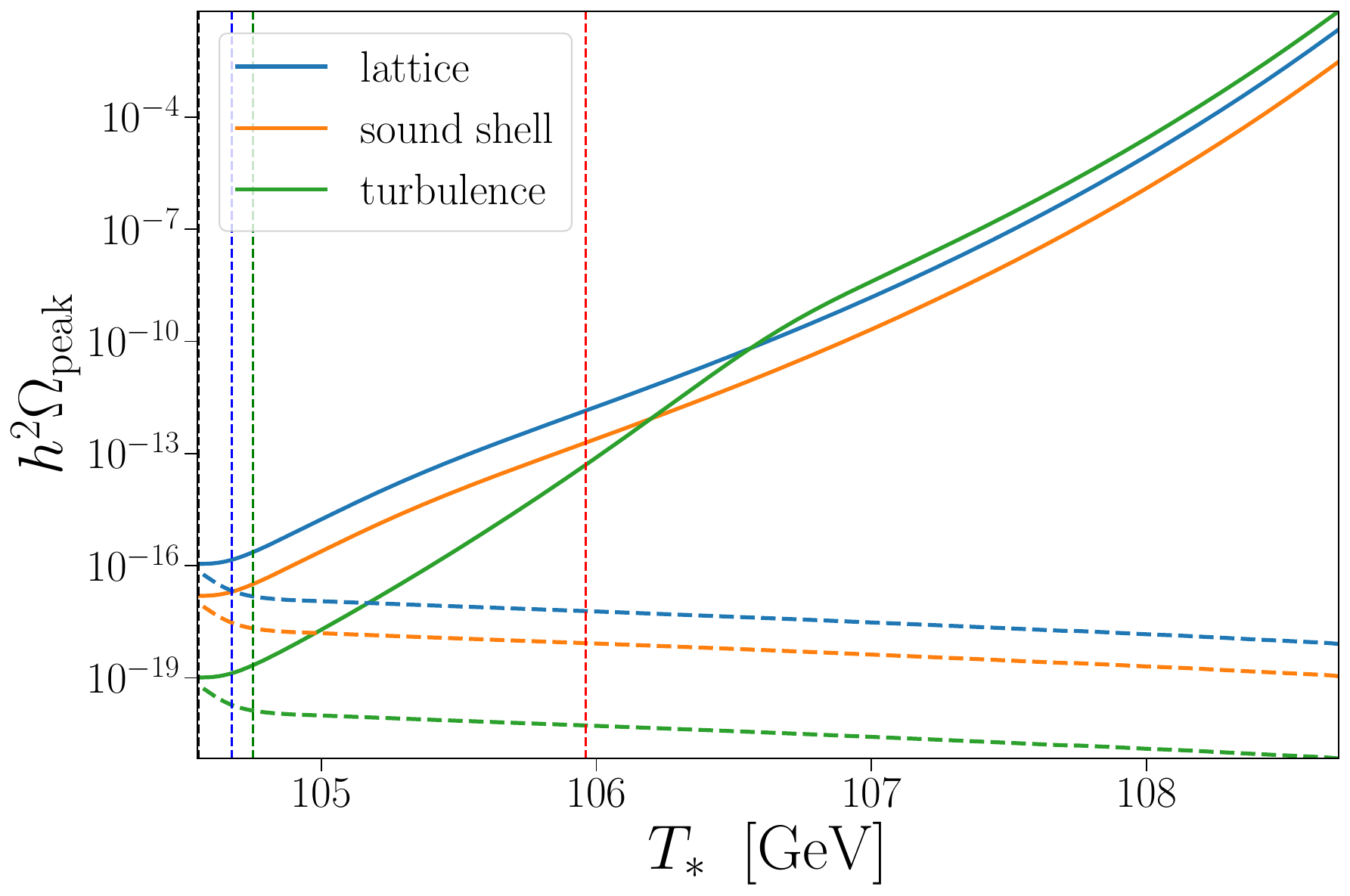}\\
	\includegraphics[width=0.95\linewidth]{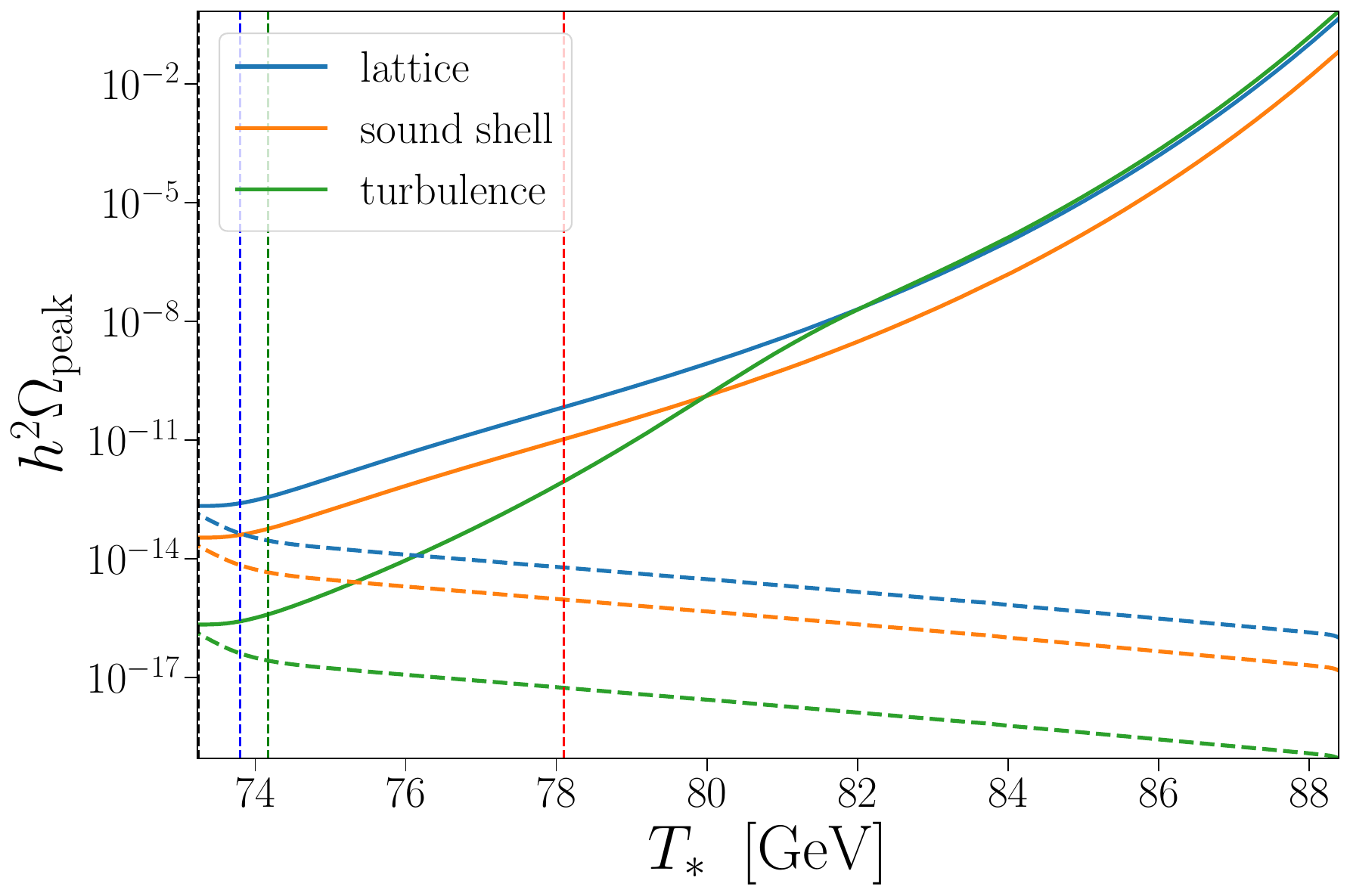}\\
	\includegraphics[width=0.95\linewidth]{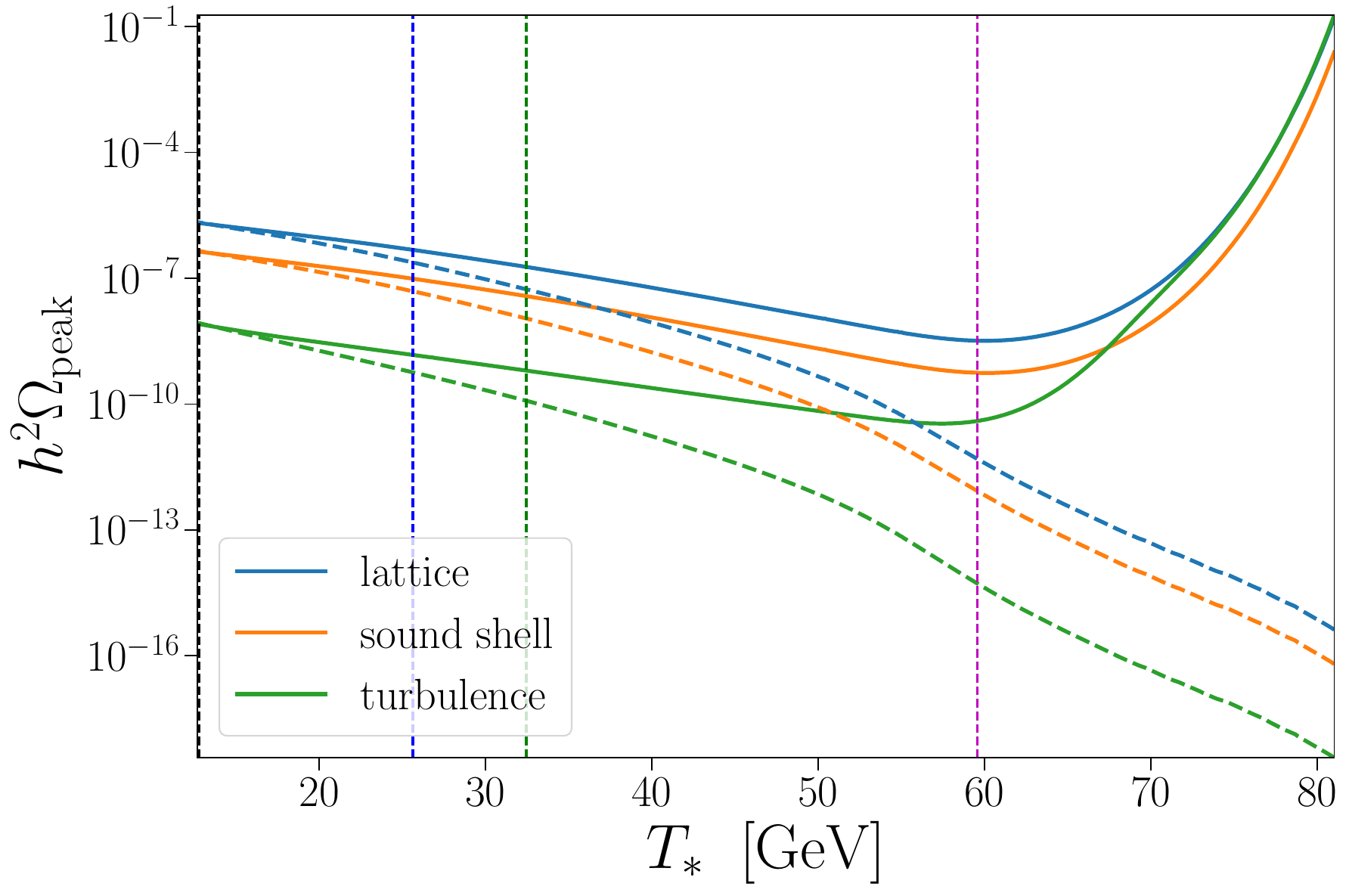}\\
	\includegraphics[width=0.95\linewidth]{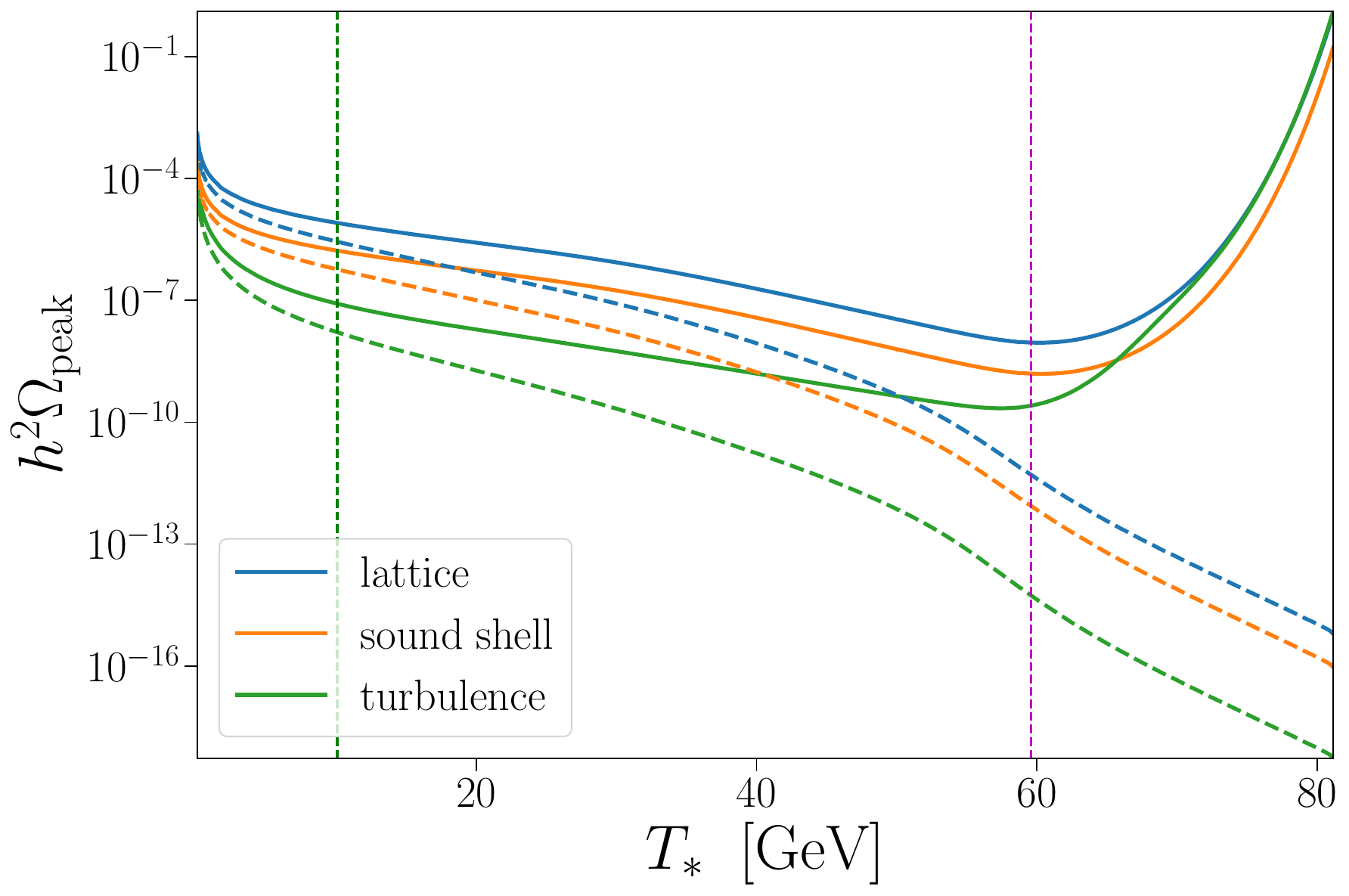}
	\caption{Peak GW amplitude as a function of transition temperature. See the caption of \cref{fig:SNR-vs-T} for further details.}
	\label{fig:Omega-peak-vs-T}
\end{figure}

\begin{figure}
	\includegraphics[width=0.95\linewidth]{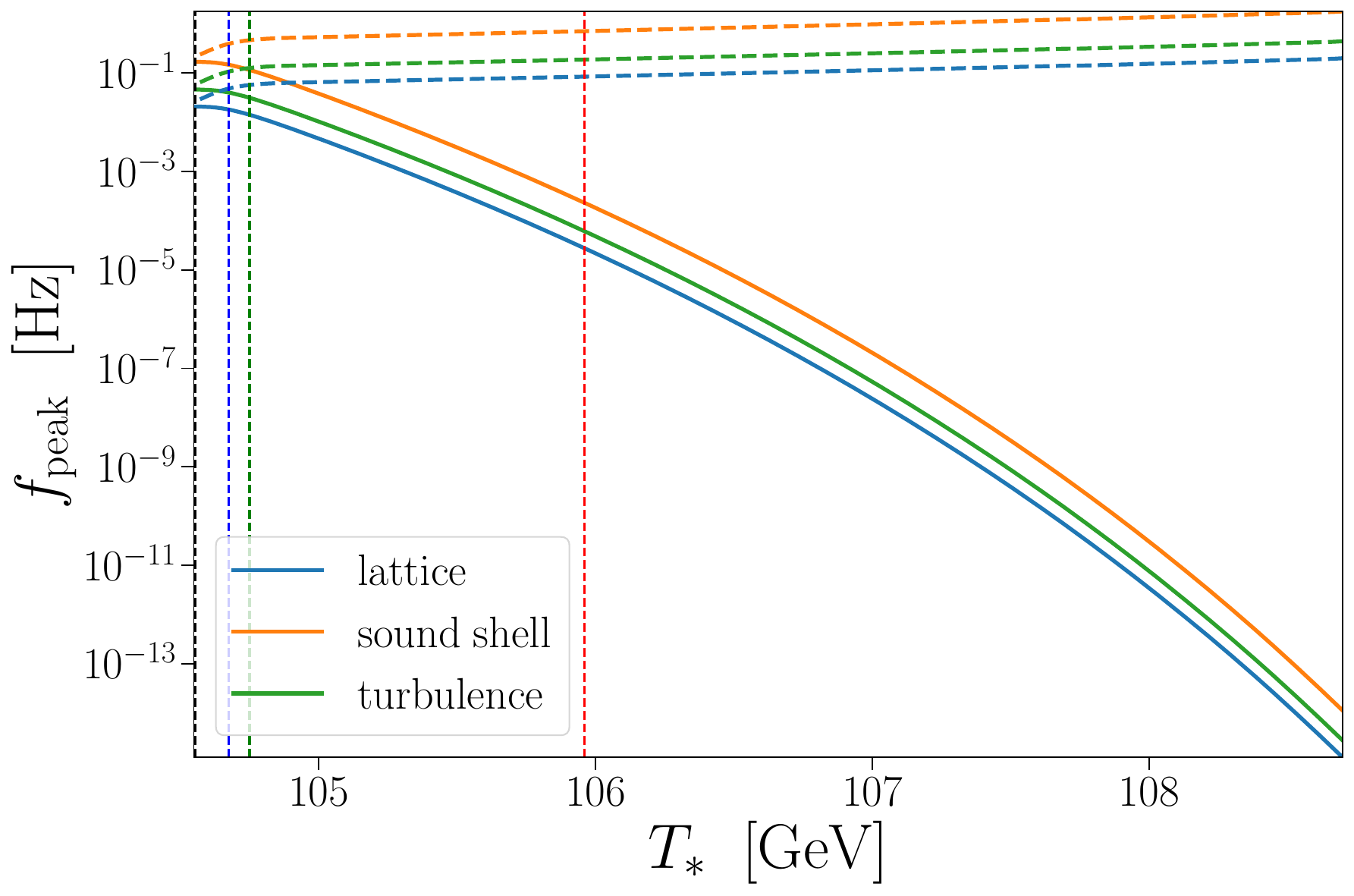}\\
	\includegraphics[width=0.95\linewidth]{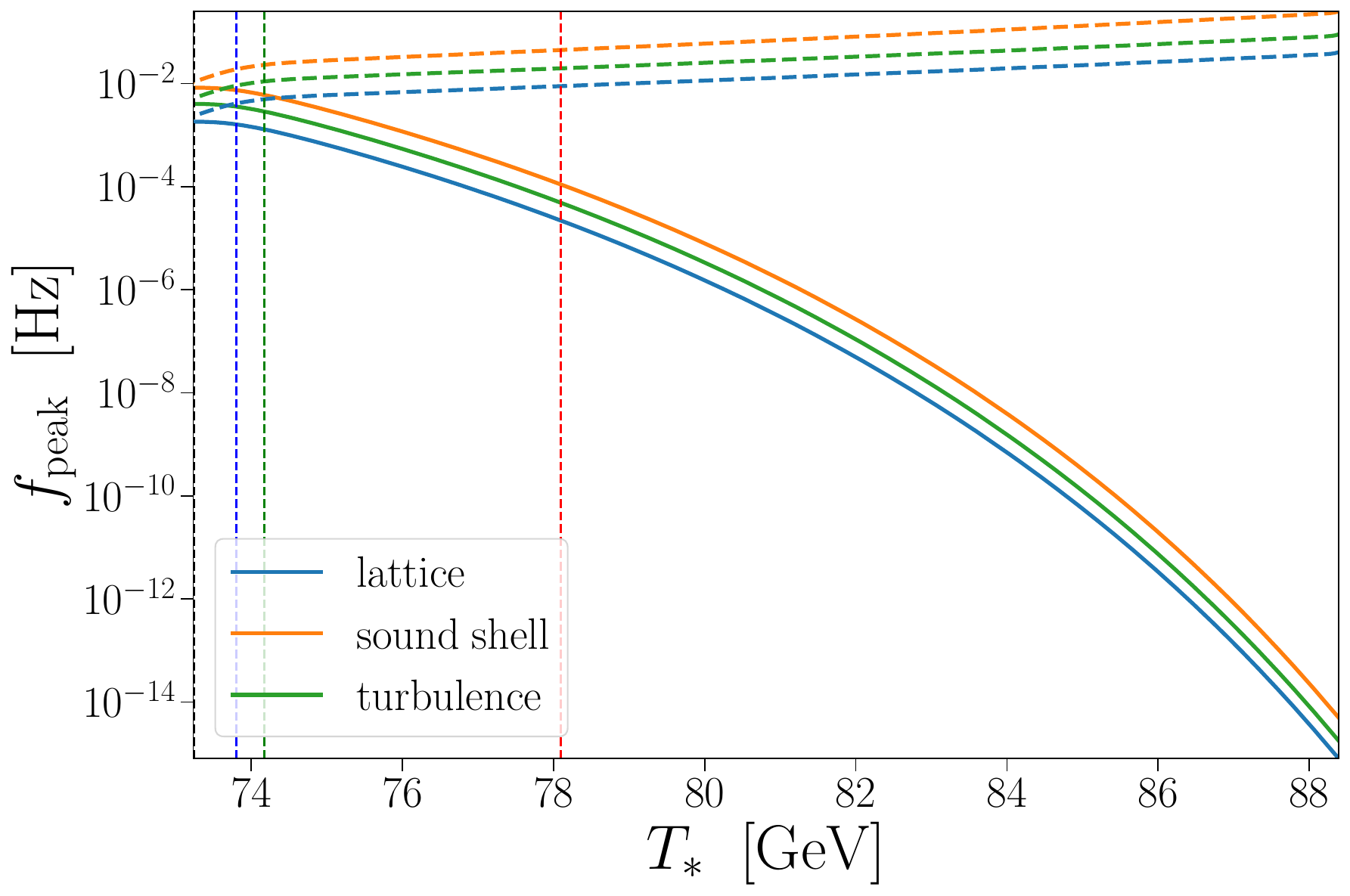}\\
	\includegraphics[width=0.95\linewidth]{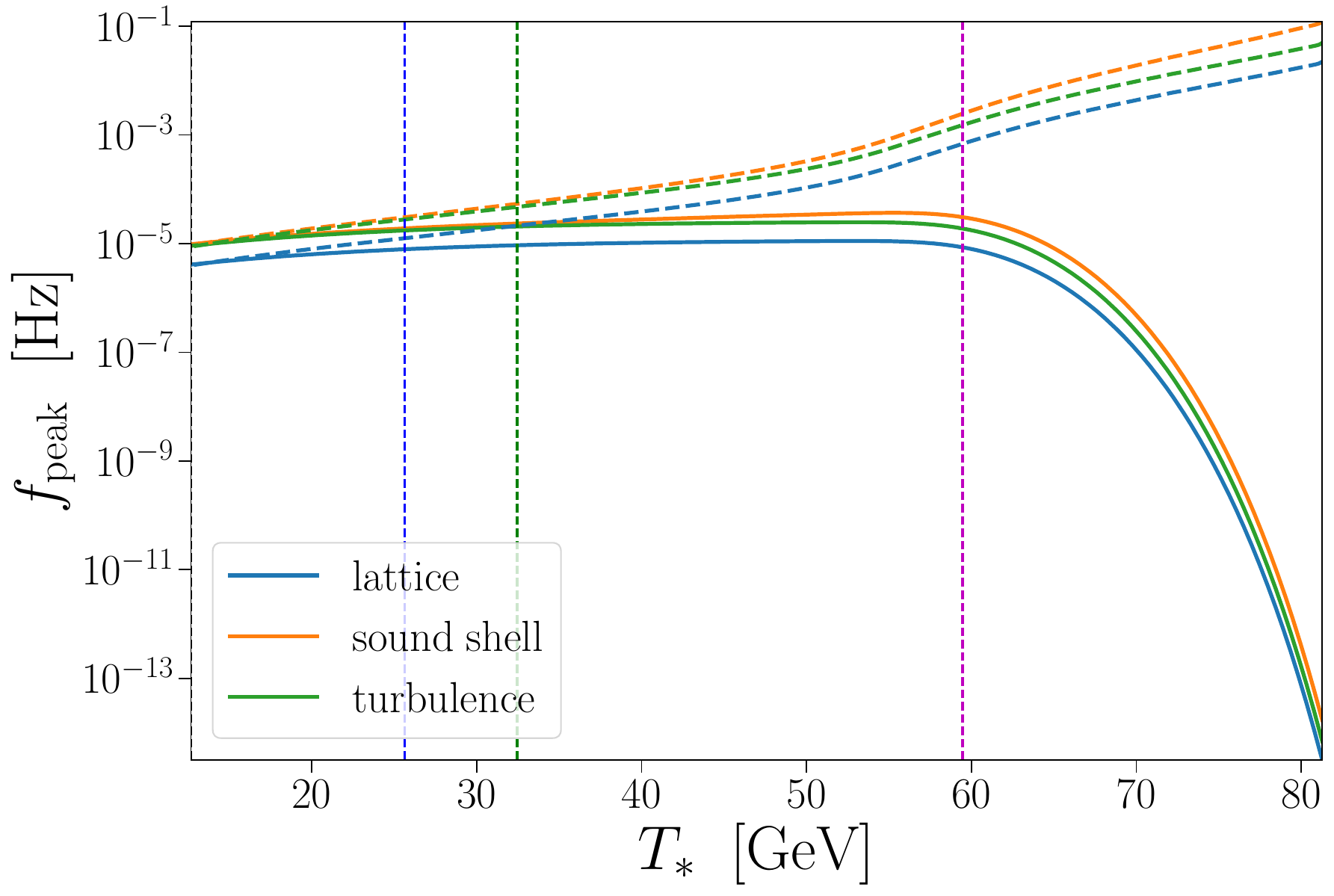}\\
	\includegraphics[width=0.95\linewidth]{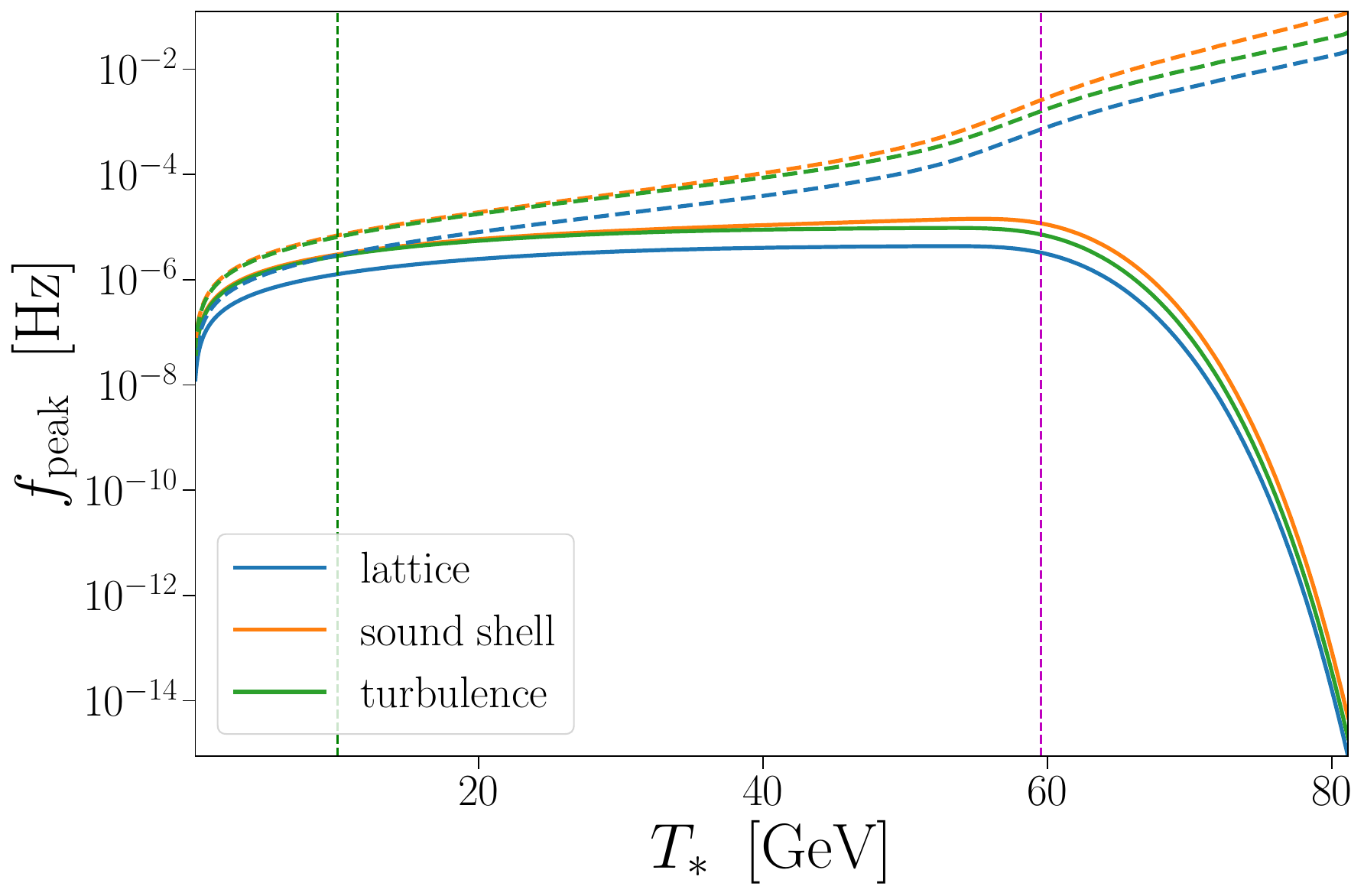}
	\caption{Peak GW frequency as a function of transition temperature. See the caption of \cref{fig:SNR-vs-T} for further details.}
	\label{fig:f-peak-vs-T}
\end{figure}

There are two sets of curves --- solid and dashed --- which have starkly different forms in the temperature domain. The solid curves use $L_* = \bubsep$ while the dashed curves use $L_* = \bubrad$, with everything else in the treatment being the same between the two sets of curves. The most immediate difference between the two sets is that the SNR increases with $T_*$ when using $\bubsep$, and decreases with $T_*$ when using $\bubrad$. In \cref{fig:Omega-peak-vs-T}(a,b) we see that the peak amplitude of GWs follows a similar behaviour: the amplitude increases (decreases) with $T_*$ when using $\bubsep$ ($\bubrad$). Inversely, in \cref{fig:f-peak-vs-T}(a,b) we see that the peak frequency of GWs decreases with $T_*$ when using $\bubsep$, and increases considerably slower with $T_*$ when using $\bubrad$.

\begin{figure}
	\centering
	\includegraphics[width=0.99\linewidth]{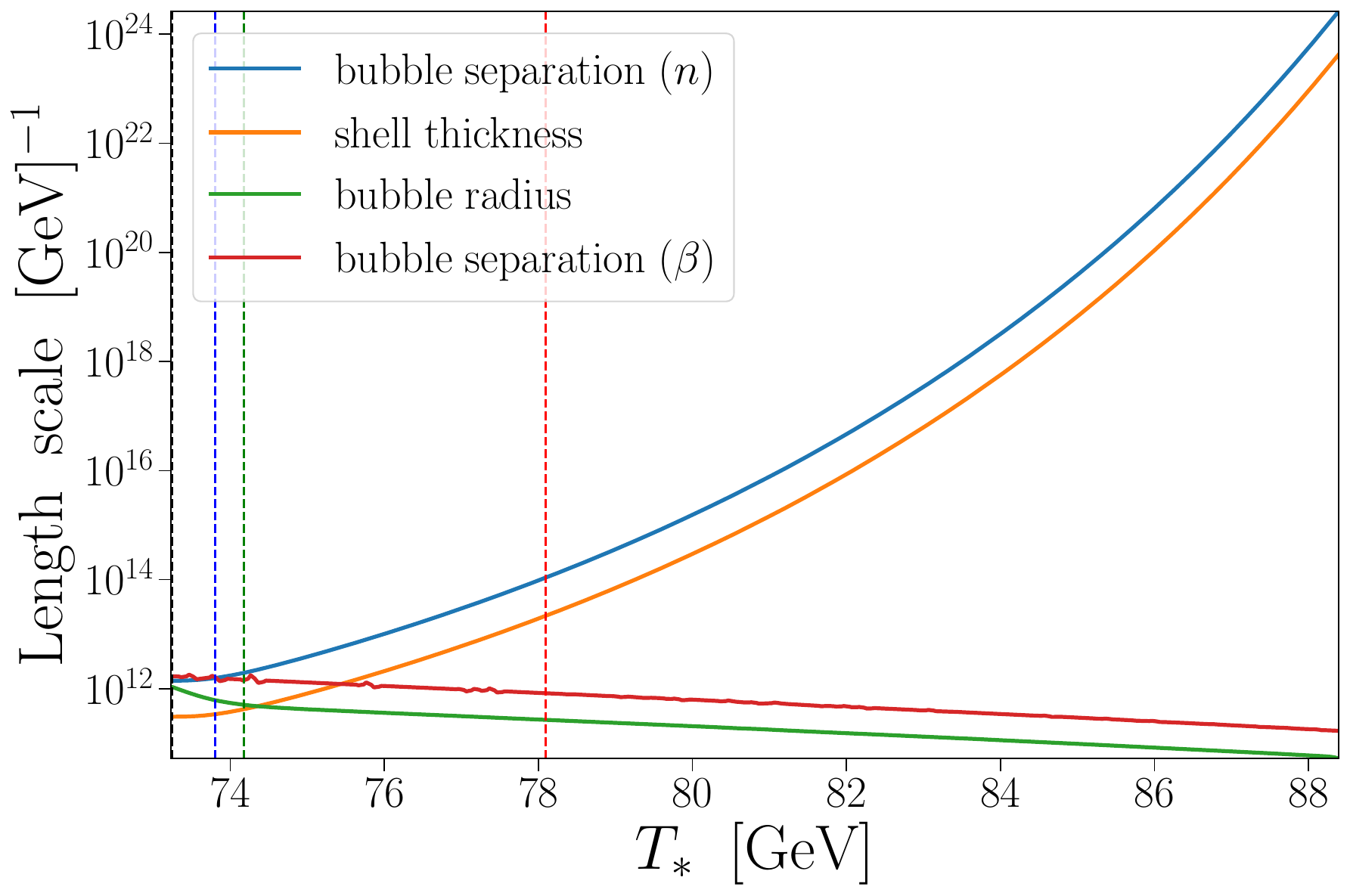}
	\includegraphics[width=0.99\linewidth]{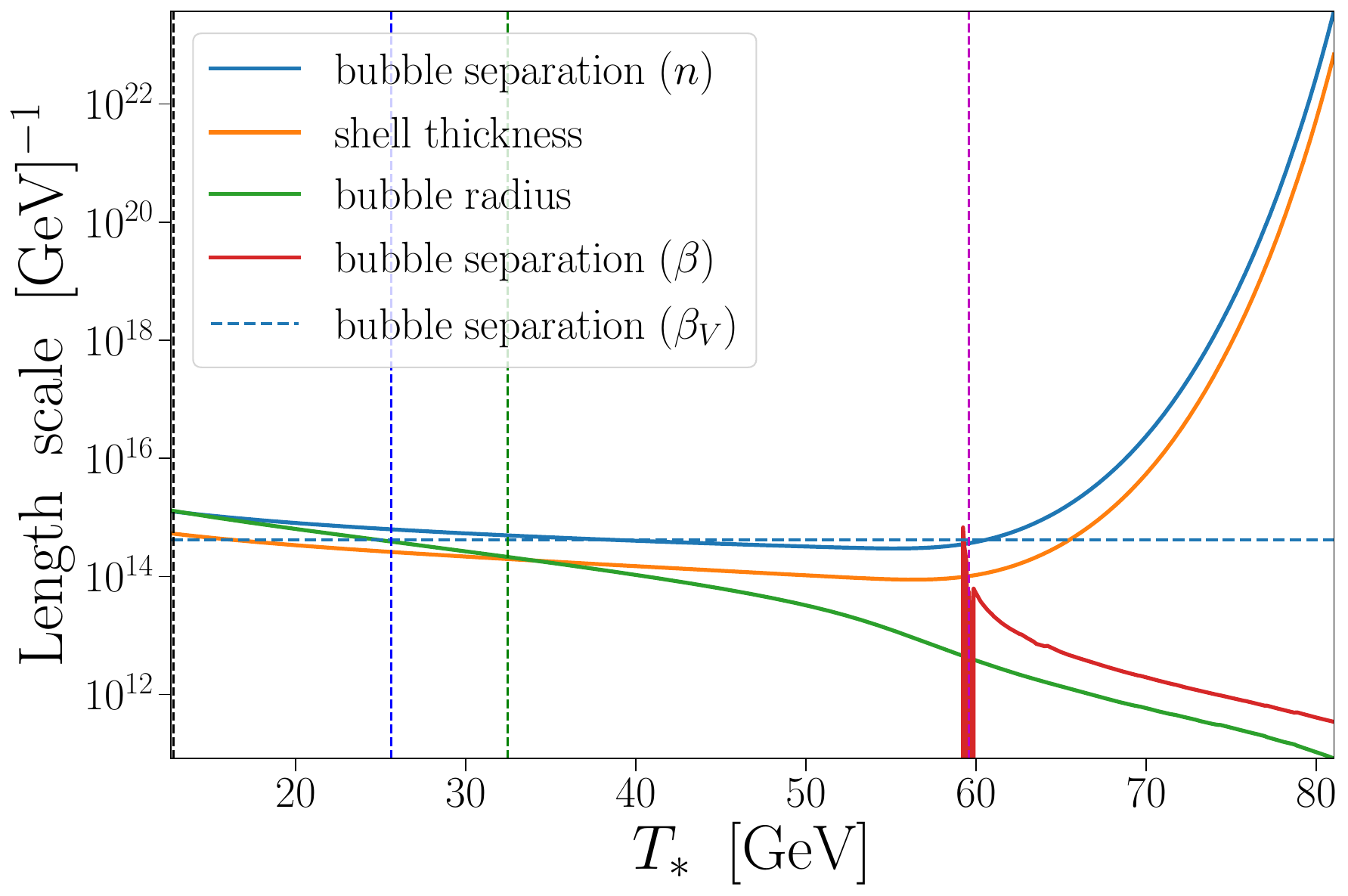}
	\caption{Characteristic length scale as a function of transition temperature. From top to bottom: (a) BP2, (b) BP3. The qualitative features of BP1 and BP4 are respectively very similar to those of BP2 and BP3, although $\bubsep$ and $\bubrad$ increase rapidly near $T_f$ in BP4. The vertical dashed lines correspond to key temperatures: $T_\Gamma$ (magenta), $T_n$ (red), $T_p$ (green), $T_e$ (blue) and $T_f$ (black). Completion occurs at the left border of each plot.}
	\label{fig:length-scale-vs-T}
\end{figure}

These observations can be easily explained by investigating the behaviour of $\bubsep$ and $\bubrad$ as function of $T_*$ (see \cref{fig:length-scale-vs-T}). In fact, we find that the dominant thermal parameter when varying $T_*$ is $L_*$, not $K$. In \cref{fig:length-scale-vs-T}(a) we plot choices of the length scale as a function of $T_*$ for BP2 (intermediate supercooling). The mean bubble separation is large near the start of the phase transition (at higher $T_*$) because there are few bubbles so their separation is large. The separation decreases over the course of the phase transition (with decreasing $T_*$) because new bubbles are nucleated. The mean bubble radius, on the other hand, begins very small because the first bubbles to nucleate have not had time to grow significantly. As the phase transition continues, pre-existing bubbles grow, but more small bubbles are nucleated, suppressing an increase in the mean radius. Thus, the mean bubble radius increases over time (i.e.\ with decreasing $T_*$) but varies less than the mean bubble separation. We also see that the mean bubble separation estimated using $\betaR$ actually emulates the mean bubble radius. This is unsurprising, because $\bubsep$ is supposedly inversely proportional to $\betaR$, and $\betaR$ is much higher at the start of a phase transition with the bounce action diverging at $T_c$. Thus, $\bubsep$ estimated using $\betaR$ is small at high $T_*$, in line with $\bubrad$, whereas the true $\bubsep$ is large at high $T_*$.

The behaviour of $\bubsep$ in BP3 (see \cref{fig:length-scale-vs-T}(b)) is more complicated due to strong supercooling. The expansion of space dilutes the bubble number density and increases the separation between bubbles. Additionally, bubble nucleation is negligible well below $T_\Gamma$ so new bubbles are not nucleated to reduce the mean separation. With even stronger supercooling in BP4 (not shown), $\bubsep$ begins to increase rapidly as $T_*$ drops below $T_p$. We also see that $\betaR$ cannot be used to estimate $\bubsep$ in BP3 (at least below $T_\Gamma$). However, one can instead use $\betaV$ under the Gaussian nucleation rate approximation, which is seen to reproduce both $\bubsep$ and $\bubrad$ quite well at $T_p$ in this example.

Now that the temperature scaling of the length scales is clear, we can return to effects on the GW signal. First, the peak frequency for all sources is inversely proportional to $L_*$ and is largely unaffected by any other thermal parameters. Only the frequency corresponding to the sound shell thickness scale (in the sound shell model) is directly affected by the hydrodynamic parameters $v_w$ and $c_s$ (and indirectly affected by $K$). The two key frequencies in the sound shell model are less separated with increased supercooling due to thickening of the sound shell. Otherwise, the behaviour of the peak frequencies in \cref{fig:f-peak-vs-T} can be explained purely by the behaviour of the length scales in \cref{fig:length-scale-vs-T}. If one uses $\bubrad$, the change in frequency with $T_*$ is milder than when using $\bubsep$. In general, stronger supercooling lowers the peak frequency at $T_p$.

Next, the peak amplitude for all sources is proportional to $L_*$. However, the amplitude also depends on $K$ and $c_s$, as well as $v_w$ indirectly through $K$. Nevertheless, $L_*$ typically has a dominant effect on the amplitude. In the absence of strong supercooling, $\bubsep$ changes considerably with $T_*$ while $\bubrad$ does not. Yet, $K$ and the other hydrodynamic parameters change very little, so $L_*$ still has a dominant effect even when using $\bubrad$. With strong supercooling, $K$ and the other hydrodynamic parameters can vary considerably between $T_\Gamma$ and $T_p$. So too can $\bubrad$, while $\bubsep$ remains approximately constant, and is in fact minimised near $T_\Gamma$. The peak amplitude increases significantly between $T_\Gamma$ and $T_p$ in BP3 when using $L_* = \bubsep$ due to the large increase in $K$ over that temperature interval. However, the peak amplitude increases rapidly below $T_p$ in BP4 not because of $K$ (which is roughly unity even at $T_p$), but because of the expansion of space causing a rapid increase in $L_*$.%
\footnote{This also causes a rapid decrease in the peak frequency at low temperature, consistent with the findings in Ref.~\cite{Athron:2023mer}.}
These are all generic features of strongly supercooled phase transitions so the results and analysis presented here should apply to other BPs and other models.

Combining the peak amplitudes and frequencies of the GW sources, one can then compare the GW signal to the sensitivity of a GW detector to obtain the SNR. We consider LISA in this study, but in principle the SNR at any detector could be calculated. Although we now have a clear picture of the behaviour of the peak amplitude and frequency, the behaviour of the SNR is complicated by the sensitivity window of LISA. The SNR is enhanced when the peak frequency matches the frequency range where LISA is most sensitive; that is, near $f_\text{LISA} \sim 10^{-3}$ Hz. If by varying $T_*$ one would obtain a higher peak amplitude but shift the peak frequency further from LISA's optimal frequency range, the SNR could decrease. Thus, investigating the peak amplitude or peak frequency in isolation will not give a clear indication of detectability.

\begin{figure}
	\centering
	\includegraphics[width=0.99\linewidth]{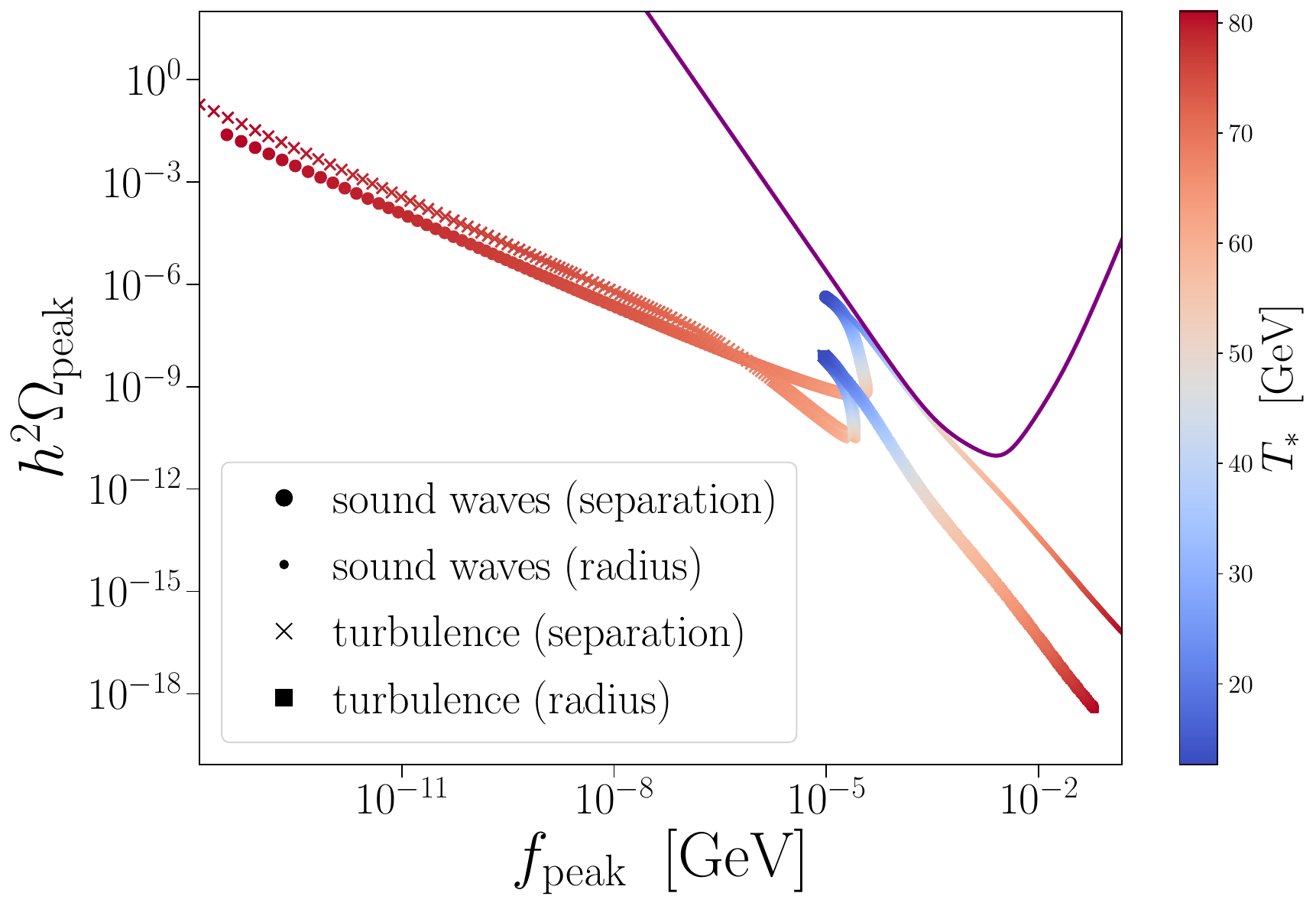}
	\caption{The peak amplitude and frequency of the GW signal for BP3 as a function of temperature. Here we show only the sound shell model for the sound wave source. The noise curve for LISA is shown in purple.}
	\label{fig:gw-peak-scatter}
\end{figure}

In \cref{fig:gw-peak-scatter} we plot the peak of the GW signal in the amplitude-frequency plane as a function of $T_*$ for BP3 to provide further insight into these competing effects.
We see that when using $\bubsep$ (the left-most curves) for a strongly supercooled phase transition, as the temperature initially decreases from high temperatures (indicated by red), the peak frequency (amplitude) increases (decreases), until a reversal occurs at the lower temperature $T_\Gamma$.  However, between $T_\Gamma$ and $T_p$ the amplitude increases faster than the frequency decreases, increasing the SNR at LISA. Meanwhile, if one uses $\bubrad$ for a strongly supercooled phase transition, the peak frequency (amplitude) decreases (increases) with decreasing $T_*$. In the example of BP3, the peak of the GW signal slides across the boundary of LISA's sensitivity curve, leading to an almost constant SNR between $T_\Gamma$ and $T_f$. One can imagine that a slightly different BP could alter the GW peak scaling slightly, leading to a substantially different scaling of SNR with $T_*$. Naturally, the curves for $\bubsep$ and $\bubrad$ meet near $T_f$ because the two length scales are very similar near the end of the phase transition (as was also demonstrated in Ref.~\cite{Megevand:2016lpr}).

The GW signal is formed from the sound wave and turbulence contributions, noting again that we have neglected the collision contribution. We consider one GW fit for the turbulence throughout, but we present results for two GW fits for sound waves: the sound shell model and lattice fits. First we compare the two fits for the sound wave source. Based on the SNR alone (see \cref{fig:SNR-vs-T}) we find a significant discrepancy between the two fits at $T_p$ in BP1 and BP2. The fits agree quite well for BP3 and BP4 when using $\bubsep$ but this is a coincidence due to LISA's sensitivity window. Looking at the peak amplitudes and frequencies separately for BP3 and BP4 (see \cref{fig:Omega-peak-vs-T}(c,d) and \cref{fig:f-peak-vs-T}(c,d)), we see that the predicted GW signals are still different. When using $\bubrad$ instead, the SNR of the sound shell model is consistently smaller in BP1 and BP2 for all $T_*$ because the peak frequency is always above LISA's optimal frequency, $f_\text{LISA}$. The situation is more complicated in BP3 and BP4 because the peak frequency crosses $f_\text{LISA}$ as $T_*$ is varied.

The ratio of peak amplitudes in the two sound wave fits is $\Omega_\text{sw}^\text{ss} / \Omega_\text{sw}^\text{lat} \approx 0.20$ for $v_w \sim 1$ and $c_{s,f} \sim 1/\sqrt{3}$, where the superscripts `ss' and `lat' denote the sound shell and lattice fits, respectively. This ratio is approximately independent of $T_*$ and is similar for all BPs. The ratio of peak frequencies is $f_\text{sw}^\text{ss} / f_\text{sw}^\text{lat} \approx 2.4$ for $v_w \sim 1$ and $c_{s,f} \sim 1/\sqrt{3}$ as in BP3, but increases to roughly $8.1$ in BP1 where $\vcj \approx 0.65$. The ratio of peak frequencies has a slight dependence on $T_*$ due to our choice $v_w = \vcj$, with $\vcj$ implicitly depending on $T_*$ through $\alpha$. The large frequency ratio in BP1 and BP2 leads to a large difference in the SNR at LISA between the two sound wave fits. The choice $v_w = \vcj$ results in a large separation in length scales --- $L_*$ and $L_* \Delta_w$ --- when $\vcj \sim c_{s,f}$, which occurs when $\alpha \ll 1$. Here, $\Delta_w = (v_w - c_{s,f}) / v_w$ is a multiplier for the sound shell thickness, and can be applied to either $\bubsep$ or $\bubrad$.

\begin{table*}[]
	\resizebox{2.0\columnwidth}{!}{
		\begin{tabular}{|l|l|l|l|l|l|l|l|l|l|l|l|}
			\hline
			Variation & \twolinepower{$h^2 \Omega_\mathrm{sw}^\mathrm{lat}$}{-17} & \twolinepower{$h^2 \Omega_\mathrm{sw}^\mathrm{ss}$}{-18} & \twolinepower{$f_\mathrm{sw}^\mathrm{lat}$}{-5} & \twolinepower{$f_\mathrm{sw}^\mathrm{ss}$}{-4} & \twolinepower{$h^2 \Omega_\mathrm{turb}$}{-20} & \twolinepower{$f_\mathrm{turb}$}{-5} & \twolinepower{$\mathrm{SNR_{lat}}$}{-5} & \twolinepower{$\mathrm{SNR_{ss}}$}{-7} & \twolinepower{$\alpha$}{-3} & \twolinepower{$\kappa$}{-2} & \twolinepower{$K$}{-4} \\
			\hline
			None				& 22.57 & 31.49 & 1422 & 1157 & 21.28 & 3150 & 156.2 & 39.60 & 11.52 & 9.900 & 11.20 \\
			\hline
			$T_* = T_e$			& 13.97 & 19.50 & 1833 & 1490 & 12.90 & 4061 & 56.44 & 11.24 & 11.57 & 9.921 & 11.27 \\
			\hline
			$T_* = T_f$			& 11.10 & 15.50 & 2080 & 1685 & 10.16 & 4607 & 33.82 & 6.105 & 11.66 & 9.955 & 11.39 \\
			\hline
			$T_* = T_n$			& 147000 & 204300 & 2.611 & 2.187 & 5448000 & 5.785 & 10230 & 5026000 & 10.74 & 9.565 & 10.09 \\
			\hline
			$\bubsep(\betaR)$  & 11.04 & 15.41 & 2062 & 1678 & 10.12 & 4567 & 34.32 & 6.216 & \matches & \matches & \matches \\
			\hline
			$K(\alpha(\theta))$ & 21.09 & 29.44 & \matches & \matches & 19.92 & \matches & 146.0 & 37.03 & 11.46 & 9.466 & 10.72 \\
			\hline
			$K(\alpha(p))$      & 1.403 & 1.957 & \matches & \matches & 1.489 & \matches & 9.711 & 2.509 & 3.590 & 5.317 & 1.902 \\
			\hline
			$K(\alpha(\rho))$   & 261.9 & 365.5 & \matches & \matches & 234.7 & \matches & 1813 & 456.2 & 35.05 & 16.39 & 55.50 \\ 
			\hline
			$\epsilon_2$		& \matches & \matches & \matches & \matches & 60.18 & \matches & 156.4 & 54.06 & \matches & \matches & \matches \\
			\hline
			$\epsilon_3$		& \matches & \matches & \matches & \matches & 1776 & \matches & 166.0 & 1035 & \matches & \matches & \matches \\
			\hline
			$\epsilon_4$		& \matches & \matches & \matches & \matches & 1787 & \matches & 166.0 & 1041 & \matches & \matches & \matches \\
			\hline
		\end{tabular}
	}
	\caption{GW predictions and hydrodynamic parameters for BP1. Each row corresponds to a different variation of our best treatment. Blank cells match the result of our best treatment (i.e.\ the top row). Frequencies are stated in units of GeV, with all other quantities being dimensionless. The scripts `ss' and `lat' respectively denote the sound shell model fit and the lattice fit for the sound wave source of GWs.}
	\label{tab:BP1}
\end{table*}

Next we compare the sound wave source to the turbulence source. In general, $\Omega_\text{turb}$ decreases faster than $\Omega_\text{sw}$ with decreasing $T_*$ when using $\bubsep$, as seen in \cref{fig:Omega-peak-vs-T}. This is because both amplitudes are proportional to the decreasing $L_*$, but $\Omega_\text{sw}$ is proportional to the increasing $K^2$ while $\Omega_\text{turb}$ is proportional to $K^{3/2}$. Thus, the fractional contribution of turbulence to the total GW signal decreases with $T_*$. However, when $K \sim 1$, as in BP4 below $T_p$, the scaling with $K$ is equivalent between the two GW sources. The comparison of the two sources does not change when instead using $\bubrad$, although the amplitudes now monotonically increase with decreasing $T_*$. There is little insight to gain when comparing the peak frequencies of the GW sources because they largely differ by a constant factor (see \cref{fig:f-peak-vs-T}). The peak frequency for the turbulence contribution is between the peak frequencies of the two sound wave fits; it is larger than that of the lattice fit and smaller than that of the sound shell model fit. However, because the sound shell thickens with supercooling (at least when choosing $v_w = \vcj$), we find that the peak frequency of turbulence closely matches the peak frequency in the sound shell model in strongly supercooled scenarios. Though, the GW fits were obtained in weak and intermediate supercooling scenarios, so their use in scenarios with strong supercooling requires extrapolation and should be interpreted with care.

Finally, one can compare the contribution to the SNR from the sound wave and turbulence sources. This information cannot be inferred from the results shown in \cref{fig:SNR-vs-T}. Instead, we will discuss the turbulence contribution --- and the impact on the SNR when increasing it --- in the next section, where we consider variations of our best treatment.

Before we proceed with that discussion we briefly consider the
meaning of the temperature dependence we have discussed here.  First,
we have seen substantial differences between predictions evaluated at
the nucleation and percolation temperatures for both slow and --- more
surprisingly --- fast transitions.  Because unit nucleation~\cite{Athron:2022mmm} can occur long
before (or after) any bubble collisions take place, the thermal
parameters evaluated at $T_n$ do not represent how they influence
the prediction of GWs sourced in the wake of bubble
collisions. Thus, evaluating GW fits at $T_n$ is not appropriate.
Instead, the percolation temperature, when defined
fundamentally in terms of connected bubbles, is a much better choice.

However, the condition $P_f(T_p)= 0.71$ we use to estimate $T_p$
is obtained from percolation simulations that do not take account of expanding spacetime. There are two competing effects in a phase transition occurring in expanding spacetime: 1) nucleation and growth of true vacuum bubbles, which reduces the physical volume of the false vacuum, $\mathcal{V}_\text{phys}$; and 2) expansion of spacetime, which increases $\mathcal{V}_\text{phys}$.
In extreme cases, expansion can
lead to $\mathcal{V}_\text{phys}$ actually increasing
such that the bubbles never meet (as discussed in \cref{sec:model}). Thus, we use a reduction in $\mathcal{V}_\text{phys}$
as an additional test of percolation and completion. In milder cases we expect expansion will merely delay percolation. This
means we expect the true percolation temperature to be at some
temperature where $P_f(T) \lesssim 0.71$.

This issue is exacerbated by the fact that the generation of
GWs is not an instantaneous process and bubble
collisions do not occur at a single temperature.  For these reasons we
regard the variation in GW predictions with temperature as an uncertainty, which has not been previously considered in the literature.

\subsection{Variations of the treatment} \label{sec:results-treatment}

\begin{table*}[]
\resizebox{2.0\columnwidth}{!}{
\begin{tabular}{|l|l|l|l|l|l|l|l|l|l|l|l|}
	\hline
	Variation & \twolinepower{$h^2 \Omega_\mathrm{sw}^\mathrm{lat}$}{-13} & \twolinepower{$h^2 \Omega_\mathrm{sw}^\mathrm{ss}$}{-14} & \twolinepower{$f_\mathrm{sw}^\mathrm{lat}$}{-5} & \twolinepower{$f_\mathrm{sw}^\mathrm{ss}$}{-4} & \twolinepower{$h^2 \Omega_\mathrm{turb}$}{-16} & \twolinepower{$f_\mathrm{turb}$}{-5} & \multicolumn{1}{|c|}{$\mathrm{SNR_{lat}}$} & \multicolumn{1}{|c|}{$\mathrm{SNR_{ss}}$} & \twolinepower{$\alpha$}{-2} & $\kappa$ & \twolinepower{$K$}{-3} \\
	\hline
	None				& 3.590 & 5.673 & 129.6 & 60.20 & 3.898 & 287.0 & 10.08 & 2.031 & 5.450 & 0.2074 & 10.64 \\
	\hline
	$T_* = T_e$ 		& 2.552 & 4.042 & 159.9 & 73.75 & 2.662 & 354.2 & 8.763 & 1.204 & 5.575 & 0.2096 & 10.99 \\
	\hline
	$T_* = T_f$ 		& 2.146 & 3.410 & 181.7 & 82.91 & 2.187 & 402.5 & 8.110 & 0.8892 & 5.771 & 0.2129 & 11.54 \\
	\hline
	$T_* = T_n$ 		& 676.5 & 1046 & 2.189 & 1.098 & 8968 & 4.849 & 1.310 & 5.142 & 4.297 & 0.1857 & 7.597 \\
	\hline
	$\bubsep(\betaR)$  	& 2.019 & 3.191 & 177.5 & 82.45  & 2.078 & 393.1 & 7.510 & 0.8449 & \matches & \matches & \matches \\
	\hline
	$K(\alpha(\theta))$ & 3.372 & 5.329 & \matches & \matches & 3.676 & \matches & 9.469 & 1.908 & 5.362 & 0.2011 & 10.23 \\
	\hline
	$K(\alpha(p))$ 		& 1.428 & 2.256 & \matches & \matches & 1.649 & \matches & 4.010 & 0.8081 & 3.698 & 0.1682 & 5.997 \\
	\hline
	$K(\alpha(\rho))$ 	& 14.45 & 22.84 & \matches & \matches & 14.61 & \matches & 40.59 & 8.172 & 10.35 & 0.2736 & 25.68 \\
	\hline
	$\epsilon_2$  		& \matches & \matches & \matches & \matches & 11.03 & \matches & 10.11 & 2.064 & \matches & \matches & \matches \\
	\hline
	$\epsilon_3$  		& \matches & \matches & \matches & \matches & 290.2 & \matches & 11.21 & 3.406 & \matches & \matches & \matches \\
	\hline
	$\epsilon_4$  		& \matches & \matches & \matches & \matches & 301.7 & \matches & 11.26 & 3.462 & \matches & \matches & \matches \\
	\hline
\end{tabular}
}
\caption{The same as \cref{tab:BP1} but for BP2.}
\label{tab:BP2}
\end{table*}

\begin{table*}[]
\resizebox{2.0\columnwidth}{!}{
\begin{tabular}{|l|l|l|l|l|l|l|l|l|l|l|l|}
	\hline
	Variation & \twolinepower{$h^2 \Omega_\mathrm{sw}^\mathrm{lat}$}{-7} & \twolinepower{$h^2 \Omega_\mathrm{sw}^\mathrm{ss}$}{-8} & \twolinepower{$f_\mathrm{sw}^\mathrm{lat}$}{-6} & \twolinepower{$f_\mathrm{sw}^\mathrm{ss}$}{-6} & \twolinepower{$h^2 \Omega_\mathrm{turb}$}{-10} & \twolinepower{$f_\mathrm{turb}$}{-6} & \multicolumn{1}{|c|}{$\mathrm{SNR_{lat}}$} & \multicolumn{1}{|c|}{$\mathrm{SNR_{ss}}$} & $\alpha$ & $\kappa$ & $K$ \\
	\hline
	None				& 1.861 & 3.748 & 9.345 & 23.48 & 6.348 & 20.70 & 249.6 & 307.7 & 1.651 & 0.7175 & 0.4536 \\
	\hline
	$T_* = T_e$			& 4.318 & 8.872 & 7.908 & 19.12 & 14.74 & 17.52 & 443.7 & 498.2 & 4.257 & 0.8422 & 0.6950 \\
	\hline
	$T_* = T_f$			& 17.04 & 35.42 & 4.111 & 9.722 & 81.84 & 9.106 & 864.5 & 876.4 & 71.06 & 0.9831 & 0.9803 \\
	\hline
	$\bubsep(\betaV)$	& 1.193 & 2.402 & 12.80 & 32.17 & 3.394 & 28.36 & 222.6 & 356.9 & \matches & \matches & \matches \\
	\hline
	$K(\alpha(\theta))$	& 1.819 & 3.663 & \matches & \matches & 6.227 & \matches & 244.9 & 301.5 & 1.605 & 0.7269 & 0.4478 \\
	\hline
	$K(\alpha(p))$		& 1.768 & 3.560 & \matches & \matches & 6.083 & \matches & 239.2 & 294.2 & 1.564 & 0.7269 & 0.4409 \\
	\hline
	$K(\alpha(\rho))$	& 1.967 & 3.962 & \matches & \matches & 6.646 & \matches & 261.4 & 323.0 & 1.728 & 0.7383 & 0.4677 \\
	\hline
	$\epsilon_2$		& \matches & \matches & \matches & \matches & 17.95 & \matches & 700.0 & 742.2 & \matches & \matches & \matches \\
	\hline
	$\epsilon_3$		& \matches & \matches & \matches & \matches & 0 & \matches & 18.36 & 130.9 & \matches & \matches & \matches \\
	\hline
	$\epsilon_4$		& \matches & \matches & \matches & \matches & 288.4 & \matches & 11210 & 11230 & \matches & \matches& \matches \\
	\hline
\end{tabular}
}
\caption{The same as \cref{tab:BP1} but for BP3. There is no row for $T_* = T_n$ because there is no nucleation temperature for BP3. This time there is a row for $\bubsep(\betaV)$ instead of $\bubsep(\betaR)$ because the bubble nucleation rate is Gaussian rather than exponential. In fact, $\betaR$ is negative and leads to invalid predictions.}
\label{tab:BP3}
\end{table*}

We now discuss the impact of individual variations to our best treatment for GW prediction. These variations involve estimating $\bubsep$ using $\betaR$ and $\betaV$, estimating $K$ using other hydrodynamic quantities, and changing the efficiency coefficient for turbulence, as discussed in \cref{sec:hydro-params}. The numerical results are stated in \cref{tab:BP1,tab:BP2,tab:BP3} for BP1-3. We do not consider BP4 here because the phase transition does not complete; besides the results should qualitatively match those of BP3. Note that studies typically do not vary from our best treatment by one small change. Usually many approximations are made for all thermal parameters used in GW predictions. Our investigation here does not encompass such treatments; instead we point the reader to Ref.~\cite{Guo:2021qcq} where they compare low and high diligence treatments. However, one cannot easily determine from their results the effects of individual variations to indicate whether an approximation is appropriate.

First, we briefly discuss the impact of the varying the transition temperature, which is otherwise treated in more detail in \cref{sec:results-temperature}. The two main parameters affecting the GW predictions are $K$ and $L_*$. We see that $K$ changes by at most a factor of a few between $T_n$ and $T_f$ even in the strongly supercooled scenario, BP3.%
\footnote{Evaluating the GW signal at $T_f$ (defined by $P_f(T_f) = 0.01$) is not a standard treatment. We show this variation to demonstrate the limiting behaviour of quantities near the end of the phase transition.}
Yet the peak amplitudes and frequencies change by several orders of magnitude. This is because $\bubsep$ changes by several orders of magnitude between $T_n$ and $T_f$. Whether the SNR is higher or lower for some choice of $T_*$ depends on where the peak frequency lies with respect to LISA's peak sensitivity, $f_\text{LISA}$. Because of this, there is no consistent trend in the effect of $T_*$ on the SNR across the BPs, even though there is a consistent trend in the peak amplitudes and frequencies.

Next, we find that using $\betaR(T_p)$ to estimate $\bubsep(T_p)$ results in roughly a factor of two error in peak amplitudes and frequencies in BP1 and BP2. A similar error is present when using $\betaV$ to estimate $\bubsep(T_p)$ in BP3. However, it is common practice to evaluate $\betaR$ at $T_n$ rather than at $T_p$, which introduces a larger error as seen in \cref{fig:length-scale-vs-T}(a). Yet using $\betaR(T_n)$ is more appropriate than using $\bubsep(T_n)$ simply because the bubble number density changes faster than $\betaR$ between $T_n$ and $T_p$. We do not consider the variation $L_* = \bubrad$ here because GW fits are derived in terms of $\bubsep$ rather than $\bubrad$. An appropriate mapping would need to be applied to use $\bubrad$ in the fits, such as multiplying $L_*$ by an unknown constant factor in the fits.

Varying the hydrodynamic quantity $x$ in \cref{Eq:Kalpha} has a significant impact on the prediction of $K$ in BP1 and BP2. The effect is considerably smaller in BP3. This can be understood as follows. The pressure difference $\Delta p$ and energy density difference $\Delta \rho$ are starkly different at high temperature, with $\Delta p = 0$ and $\Delta \rho \neq 0$ at $T_c$. We always have $\alpha_p < \alpha_\theta < \alpha_\rho$~\cite{Giese:2020rtr}. Using the pressure difference underestimates $K$, while using the energy density difference overestimates $K$. Our results match the findings of Refs.~\cite{Giese:2020rtr,Giese:2020znk}.\texttt{} With increased supercooling (i.e.\ at lower temperature), the energy density approaches the pressure such that $\alpha_p \approx \alpha_\rho$, and $c_{s,t}^2 \approx 1/3$ such that $\pt \approx \theta$. Thus, for strong supercooling we find that all methods to estimate $K$ lead to similar results, while significant discrepancies arise for weak and intermediate supercooling.

Lastly, we consider the impact of varying the turbulence efficiency coefficient, $\kappa_\text{turb}$, through variation of $\epsilon$ (see \cref{Eq:epsilon-turb}). Increasing $\kappa_\text{turb}$ can have a large impact on the SNR, particularly if the peak frequency of turbulence better matches the detector's sensitivity window than the peak frequency of sound waves does. The variations $\epsilon_3$ and $\epsilon_4$ increase the amplitude of the turbulence source by two orders of magnitude because $\epsilon$ approaches unity, and $(1/0.05)^{3/2} \approx 90$. However, $\epsilon_3$ predicts zero turbulence in BP3 because $H(T_*) \tau_\text{sw} > 1$. Increasing the turbulence contribution increases the SNR significantly in BP1 when using the sound shell model but has little effect when using the lattice fit for sound waves. The effect is small in BP2 with up to a 50\% increase in SNR when using the sound shell model. The effect is significant in BP3 when using either sound wave fit.

\section{Discussion}
\label{sec:discussion}
In this study we have investigated several ambiguities and approximations made in predictions of GWs from cosmological phase transitions. We considered each approximation in isolation to provide a clear indication of their individual effects on the GW signal. We recommend our results be used in conjunction with the results of Ref.~\cite{Guo:2021qcq} to determine whether a particular set of approximations can lead to reliable GW predictions. Alternatively, one could use our best treatment described in \cref{sec:hydro-params} if feasible, and even improve on it with a proper treatment of hydrodynamic profile around bubble walls and a method for estimating friction on the bubble wall.

To our knowledge, our investigation is the first to explicitly determine the effect of varying the transition temperature, $T_*$. We note that our investigation is fundamentally different from studies that vary thermal parameters (including $T_*$) separately, treating them as independent quantities. We account for the implicit interdependence of all thermal parameters.

The correct choice of the transition temperature is still unknown because the hydrodynamic simulations from which GW fits are obtained hold the temperature fixed. In fact, evaluating GW predictions at a single temperature may fall out of favour once modelling of GW production is improved further. We have demonstrated that using the current set of thermal parameters (in particular $\bubsep$), the GW signal can change by several orders of magnitude between commonly chosen transition temperatures: $T_n$ and $T_p$.  If a more appropriate choice of transition temperature deviates from $T_p$, then new GW predictions would significantly differ from those obtained using the current best treatments which use $T_* = T_p$.

We argued in \cref{sec:results-temperature} that evaluating the GW signal at temperatures above $T_n$ is not meaningful because bubble collisions would not have occurred to source GWs at that stage in the phase transition. This same reasoning can also be used to discard $T_n$ as a reasonable transition temperature. The only case where the nucleation temperature reflects a time when collisions are occurring is in some strongly supercooled phase transitions --- where in extreme cases $T_n \sim T_p$, counter-intuitively~\cite{Athron:2022mmm}. However, using $T_n$ in strongly supercooled phase transitions is not recommended. For one, it decouples from the progress of the phase transition, so it does not represent a consistent stage in the phase transition. Further, the existence of a nucleation temperature does not indicate whether a phase transition occurs or completes, as discussed in Ref.~\cite{Athron:2022mmm}. Thus, one must be careful when using $T_n$, and ensure that the phase transition is in fact weakly supercooled.

It is commonly assumed that the GW signal should be similar at $T_n$ and $T_p$ for weakly supercooled phase transitions. This is not consistent with our findings. Calculating the mean bubble separation properly (from the bubble number density) would suggest orders of magnitude difference in the GW signal between $T_n$ and $T_p$. Using the mean bubble radius or $\beta$ instead still suggests a factor of a few difference in the GW signal between $T_n$ and $T_p$. The hydrodynamic parameters like the kinetic energy fraction, however, are similar at the two temperatures.

The mean bubble radius varies much slower with temperature than the mean bubble separation. Thus, studies evaluating GWs at $T_n$ should use the mean bubble radius or $\betaR$ instead of calculating the mean bubble separation directly from the bubble number density. However, we note that if one could calculate the bubble number density, then one could calculate $T_p$ and use the recommended treatment outlined in \cref{sec:hydro-params}.

In general, we find that variations of the treatment of GW predictions can lead to sizeable deviations in the SNR and peak amplitudes and frequencies; potentially deviations of many orders of magnitude. In the context of GW predictions from cosmological phase transitions, mild deviation is of the order of 10\%, suggesting that constraints on particle physics models from GW observations will be hard to apply reliably at this stage. Nevertheless, the recent emergence of successful GW astronomy offers hope for constraining particle physics models at scales beyond the reach of particle physics experiments.
\acknowledgements
LM thanks Thomas Konstandin for assistance with numerical accuracy in calculating $\kappa_{\pt}$.  LM was supported by an Australian Government Research Training Program (RTP) Scholarship and a Monash Graduate Excellence Scholarship (MGES).   The work of PA is supported by the National Natural Science Foundation of China (NNSFC) under grant Nos. 12150610460 and 12335005 and by the supporting fund for foreign experts grant wgxz2022021L. ZX is also supported in part by NNSFC grant No. 12150610460.

\appendix

\section{Correction to the kinetic energy fraction parameterisation} \label{app:deltas}

The kinetic energy fraction is often parameterised as
\begin{equation}
	K = \frac{\kappa \alpha}{1 + \alpha} . \label{eq:K-alpha}
\end{equation}

\noindent This parameterisation introduces approximation to the fundamental definition~\cite{Hindmarsh:2019phv, Giese:2020rtr, Athron:2023xlk}
\begin{equation}
	K = \frac{\rho_\text{kin}(T_*)}{\rhotot(T_*)} , \label{eq:K}
\end{equation}
where $\rho_\text{kin}$ is the fluid kinetic energy. In the following we assume $\rho$ and $p$ are renormalised such that the ground state energy density vanishes. In this case, $\rhotot = \rho_f$.

The inexact nature of \cref{eq:K-alpha} was demonstrated in appendix B.2 of Ref.~\cite{Hindmarsh:2019phv} and implied in Ref.~\cite{Giese:2020rtr} (seen by comparing methods M2 and M3). A correction $\delta$ can be applied such that~\cite{Hindmarsh:2019phv}

\vspace{-0.25cm}

\begin{equation}
	K = \frac{\kappa \alpha}{1 + \alpha + \delta} . \label{eq:K-alpha-delta}
\end{equation}
One can solve for $\delta$ by equating \cref{eq:K} and \cref{eq:K-alpha-delta}. If $\alpha$ is calculated using the trace anomaly
\begin{equation}
	\theta = \frac14 \! \left(\rho - 3p \right)
\end{equation}
as in Ref.~\cite{Hindmarsh:2019phv}, one finds
\begin{equation}
	\delta = \frac{\theta_t}{3 w_f} .
\end{equation}
If $\alpha$ is calculated using the pseudotrace~\cite{Giese:2020rtr}
\begin{equation}
	\pt = \frac14 \! \left(\rho - \frac{p}{c_{s,t}^2} \right) ,
\end{equation}
which reduces to the trace anomaly if $c_{s,t}^2 = 1/3$ (e.g.\ as in the bag model), one instead finds
\begin{equation}
	\delta = \frac{4}{3w_f} \! \left(\rhotot - \Delta \pt \right) - 1 .
\end{equation}

In our benchmark points we find $\delta \ll 1 + \alpha$ such that the difference between \cref{eq:K-alpha} and \cref{eq:K-alpha-delta} is at most 1\%. Thus, we do not include such variations on the treatment of $K$ in our results.

\bibliographystyle{JHEP}
\bibliography{References}

\end{document}